\documentclass[twocolumn]{aastex62}
\usepackage{amsmath,color,textcomp,url,graphicx,subfigure}
\usepackage{listings}

\newcommand{\likelymembers}{19}
\newcommand{\age}{$89_{-7}^{+5}$}
\newcommand{\spatsize}{4.4}

\newcommand{\candidatemembers}{46}
\newcommand{\gaiacandidatemembers}{43}
\newcommand{\tmwisecandidatemembers}{3}
\newcommand{\volanscand}{6}
\newcommand{\kms}{\hbox{km\,s$^{-1}$}}

\newcommand{\msol}{$M_{\odot}$}
\newcommand{\masyr}{$\mathrm{mas}\,\mathrm{yr}^{-1}$}
\newcommand{\teff}{$T_{\rm eff}$}

\submitjournal{ApJ}

\shorttitle{The Volans-Carina Association}
\shortauthors{Gagn\'e et al.}

\begin{document}

\title{VOLANS-CARINA: A NEW 90 M\lowercase{yr} OLD STELLAR ASSOCIATION AT 85 \lowercase{pc}}

\author[0000-0002-2592-9612]{Jonathan Gagn\'e}
\affiliation{Carnegie Institution of Washington DTM, 5241 Broad Branch Road NW, Washington, DC~20015, USA}
\affiliation{NASA Sagan Fellow}
\email{jgagne@carnegiescience.edu}
\author[0000-0001-6251-0573]{Jacqueline K. Faherty}
\affiliation{Department of Astrophysics, American Museum of Natural History, Central Park West at 79th St., New York, NY 10024, USA}
\author[0000-0003-2008-1488]{Eric E. Mamajek}
\affiliation{Jet Propulsion Laboratory, California Institute of Technology, 4800 Oak Grove Drive, Pasadena, CA 91109, USA}
\affiliation{Department of Physics \& Astronomy, University of
  Rochester, Rochester, NY 14627, USA}

\begin{abstract}

We present a characterization of the new Volans-Carina Association (VCA) of stars near the Galactic plane ($b$\,$\simeq$\,-10\textdegree) at a distance of $\simeq$\,75--100\,pc, previously identified as group~30 by \cite{2017AJ....153..257O}. We compile a list of \likelymembers\ likely members from \emph{Gaia}~DR2 with spectral types B8--M2, and \candidatemembers\ additional candidate members from \emph{Gaia}~DR2, 2MASS and AllWISE with spectral types A0--M9 that require further follow-up for confirmation. We find an isochronal age of \age\,Myr based on MIST isochrones calibrated with Pleiades members. This new association of stars is slightly younger than the Pleiades, with less members but located at a closer distance, making its members $\simeq$\,3 times as bright than those of the Pleiades on average. It is located further than members of the AB~Doradus moving group which have a similar age, but it is more compact on the sky which makes it less prone to contamination from random field interlopers. Its members will be useful benchmarks to understand the fundamental properties of stars, brown dwarfs and exoplanets at $\simeq$\,90\,Myr. We also provide an updated version of the BANYAN~$\Sigma$ Bayesian classification tool that includes the Volans-Carina association.

\end{abstract}


\section{INTRODUCTION}\label{sec:intro}

The \emph{Gaia} mission \citep{2016AA...595A...1G} is reshaping our understanding of the Solar neighborhood and the Milky Way kinematics. The Data Release 1 of the \emph{Gaia} mission \citep{2016AA...595A...4L} on 2016 September 14 published precise parallaxes ($\simeq$\,0.3\,mas) and proper motions for 2 million stars in the Tycho-2 catalog \citep{2000AA...355L..27H}, more than a factor two improvement over the Hipparcos mission ($\simeq$1\,mas; \citealp{1997AA...323L..49P,2007AA...474..653V}) that was still widely used to characterize stars and young associations in the Solar neighborhood (e.g., \citealp{2004ARAA..42..685Z,2008hsf2.book..757T}). The first data release already led to the discovery of several new pairs and larger groups of co-moving stars (e.g., \citealp{2017MNRAS.472..675A,2017AJ....153..259O,2017AJ....153..257O}). The Data Release 2 of \emph{Gaia} (\emph{Gaia}~DR2 hereafter; \added{\citealt{GaiaCollaboration:2018io,Lindegren:2018gy}})\footnote{\added{See also \cite{Luri:2018eu}, \cite{Mignard:2018bj}, \cite{Babusiaux:2018di},\cite{Sartoretti:2018jm}, \cite{Soubiran:2018fz}, \cite{Cropper:2018jx}, \cite{Evans:2018cj}, \cite{Hambly:2018gr}, and \cite{Riello:2018bo} for relevant calibration.}} published $\simeq$\,1.3 billion trigonometric distances and $\simeq$\,7.2 million radial velocities on 2018 April 25 with a precision 100 times better than that of Hipparcos, and is instigating a revolution in stellar astronomy, among other fields.

\begin{figure*}
	\centering
	\includegraphics[width=0.965\textwidth]{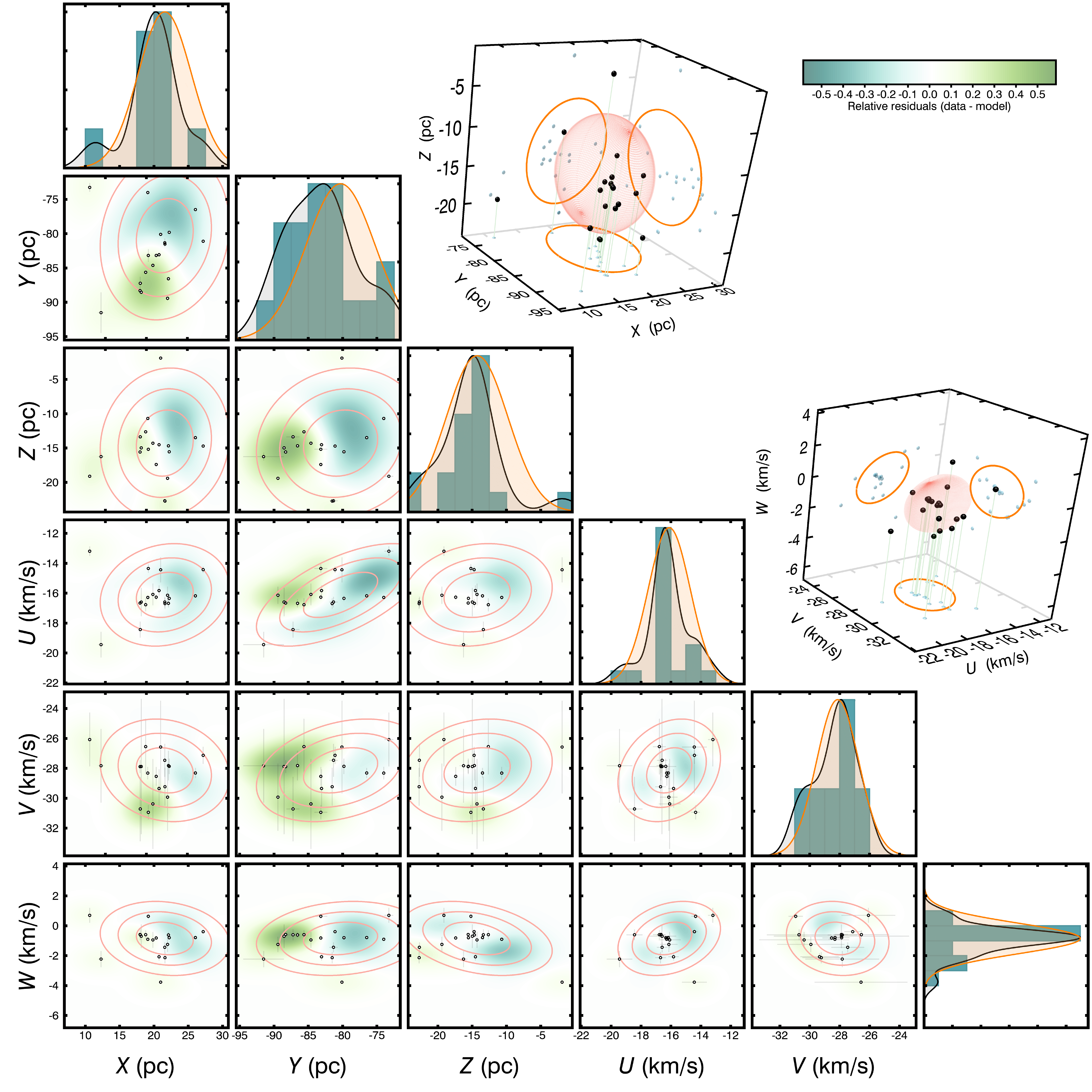}
	\caption{Projections in Galactic position $XYZ$ and space velocities $UVW$ of the Volans-Carina association members (black circles) compared to 1 to 3$\sigma$ contours of the BANYAN~$\Sigma$ multivariate Gaussian models (orange lines). Over-densities and under-densities of the data with respect to the Gaussian models are designated with blue and green shaded regions, respectively.  The 1-dimensional panels show histograms of the members (green bars) compared to the projected 1$\sigma$ model contour (orange lines) and a kernel density estimate of the members distribution (black lines). See Section~\ref{sec:model} for more detail.}
	\label{fig:oh30_params}
\end{figure*}

The new ensembles of co-moving stars identified by \cite{2017AJ....153..257O} include group~30, consisting of 8 stars at a distance of $\simeq$\,80--90\,pc with similar sky positions and proper motions. This group was further vetted by \cite{2018arXiv180409058F} where they performed a comprehensive check of the 4\,555 groups identified by \cite{2017AJ....153..257O} and found that group~30 was a new coeval association within 100\,pc with an age similar to the Pleiades ($112 \pm 5$\,Myr; \citealt{2015ApJ...813..108D}) based on its color-magnitude diagram and the X-ray properties of a single member. The reason why this group had not been identified before is likely because its members are close to the Galactic plane ($b = -10 \pm 1$\textdegree). In this paper, we use \emph{Gaia}~DR2 data to characterize this new association of stars and compile a more complete list of members and candidate members.

The 8 stars in group~30 are located in the Carina constellation, however there are already two associations of stars named after this constellation (the Carina-Near moving group; \citealp{2006ApJ...649L.115Z}; and the Carina association; \citealp{2008hsf2.book..757T}). We will show in this paper that \volanscand\ new candidate members of this group\deleted{, as well as its brightest member (the B8 supergiant c~Car)} not initially identified by \cite{2017AJ....153..257O}\deleted{,} are located in the Volans constellation. We will therefore refer to group~30 as the Volans-Carina Association ("VCA" or "Vol-Car").

\startlongtable
 \begin{deluxetable*}{lllcccccc}
\tablecolumns{9}
\tablecaption{Members of Volans-Carina.\label{tab:orig}}
 \tablehead{\colhead{VCA} & \colhead{Designation} & \colhead{Spec.} & \colhead{R.A.\tablenotemark{b}} & \colhead{Decl.\tablenotemark{b}} & \colhead{$G$} & \colhead{Trig.} & \colhead{Source\tablenotemark{c}} & \colhead{Ref.\tablenotemark{d}}
\\
 \colhead{ID} & \colhead{} & \colhead{Type\tablenotemark{a}} & \colhead{(hh:mm:ss.sss)} & \colhead{(dd:mm:ss.ss)} & \colhead{(mag)} & \colhead{dist. (pc)} & \colhead{} & \colhead{}
 }
\startdata
1 & c Car (HR~3571) & B8 II & 08:55:02.791 & -60:38:39.00 & $3.728 \pm 0.005$ & $93.8 \pm 3.0$ & 3 & 1\\
2 & HD 80563 & F3V & 09:17:34.075 & -63:23:13.88 & $8.2152 \pm 0.0004$ & $91.45 \pm 0.23$ & 1 & 2\\
3 & HD 83946 & F5V & 09:38:54.019 & -64:59:26.10 & $8.7190 \pm 0.0005$ & $90.55 \pm 0.22$ & 2 & 2\\
4 & HD 309681 & G0 & 09:25:01.560 & -64:37:30.49 & $9.026 \pm 0.001$ & $86.70 \pm 0.91$ & 2 & 3\\
5 & CD-67 852 & (K0) & 10:12:29.772 & -67:52:32.79 & $9.7674 \pm 0.0007$ & $81.92 \pm 0.15$ & 1 & --\\
6 & TYC 8933-327-1 & (K4) & 08:26:49.603 & -63:46:36.08 & $10.5379 \pm 0.0005$ & $76.45 \pm 0.60$ & 2 & --\\
7 & TYC 8953-1289-1 & (K4) & 09:39:10.366 & -66:46:14.85 & $10.7231 \pm 0.0005$ & $84.29 \pm 0.18$ & 1 & --\\
8 & TYC 8950-1447-1 & (K6) & 09:48:19.150 & -64:03:21.12 & $11.034 \pm 0.001$ & $77.17 \pm 0.14$ & 1 & --\\
9 & TYC 9210-1818-1 & (K6) & 10:08:21.890 & -67:56:39.63 & $11.2997 \pm 0.0007$ & $86.79 \pm 0.17$ & 1 & --\\
10 & 2MASS J09201987-6607418 & (K8) & 09:20:19.783 & -66:07:41.06 & $12.025 \pm 0.002$ & $87.37 \pm 0.18$ & 2 & --\\
11 & 2MASS J09345577-6500076 & K7 & 09:34:55.690 & -65:00:07.10 & $12.2903 \pm 0.0008$ & $86.84 \pm 0.25$ & 2 & 4\\
12 & 2MASS J09330015-6251207 & (M1) & 09:33:00.082 & -62:51:20.17 & $12.5210 \pm 0.0009$ & $88.57 \pm 0.18$ & 2 & --\\
13 & 2MASS J09263145-6236236\tablenotemark{e} & (M1) & 09:26:31.370 & -62:36:22.98 & $12.6322 \pm 0.0009$ & $90.07 \pm 0.92$ & 2 & --\\
14 & 2MASS J08591364-6919442 & (M2) & 08:59:13.574 & -69:19:43.48 & $12.7541 \pm 0.0004$ & $87.15 \pm 0.58$ & 2 & --\\
15 & 2MASS J10223868-5847599 & (M2) & 10:22:38.585 & -58:47:59.67 & $12.8032 \pm 0.0005$ & $82.82 \pm 0.20$ & 2 & --\\
16 & 2MASS J08591527-6918426 & (M2) & 08:59:15.199 & -69:18:41.80 & $12.812 \pm 0.001$ & $87.34 \pm 0.16$ & 2 & --\\
17 & 2MASS J09311626-6414270 & (M2) & 09:31:16.178 & -64:14:26.44 & $12.859 \pm 0.001$ & $88.08 \pm 0.26$ & 2 & --\\
18 & 2MASS J09174971-6628140\tablenotemark{e} & (M2) & 09:17:49.649 & -66:28:13.45 & $13.076 \pm 0.001$ & $94.11 \pm 0.32$ & 2 & --\\
19 & 2MASS J09203649-6309598 & (M2) & 09:20:36.427 & -63:09:59.14 & $13.111 \pm 0.002$ & $91.61 \pm 0.18$ & 2 & --\\
\enddata
\tablenotetext{a}{Spectral types between parentheses were estimated from the absolute \emph{Gaia} $G$--band magnitude with the method described in Section~\ref{sec:census}.}
\tablenotetext{b}{J2000 position at epoch 2015 from the \emph{Gaia}~DR2 catalog.}
\tablenotetext{c}{Sources for inclusion in the list of members: (1) original list from \citealt{2017AJ....153..257O}; (2) $XYZUVW$ box search in \emph{Gaia}~DR2; and (3) box search in observables without radial velocity in \emph{Gaia}~DR2.}
\tablenotetext{d}{References for spectral types.}
\tablenotetext{e}{\emph{Gaia}~DR2 absolute $G$ magnitude versus $G-G_{\rm RP}$ color is consistent with an unresolved binary.}
\tablerefs{(1)~\citealt{1994AJ....107.1556G}; (2)~\citealt{1975mcts.book.....H}; (3)~\citealt{1995AAS..110..367N}; (4)~\citealt{2006AJ....132..866R}.}
\end{deluxetable*}

An initial list of Volans-Carina members is compiled in Section~\ref{sec:ident}. In Section~\ref{sec:model}, we build a kinematic model of Volans-Carina based on the \cite{2017AJ....153..257O} compilation of members and include it in the BANYAN~$\Sigma$ Bayesian classification algorithm \citep{2018ApJ...856...23G}. We use this updated version of BANYAN~$\Sigma$ to identify additional members and candidate members based on \emph{Gaia}~DR2, 2MASS and AllWISE in Section~\ref{sec:census}. In Section~\ref{sec:discussion}, we investigate the age of the Volans-Carina association, the chromospheric activity of its members, and we build its preliminary present-day mass function. We conclude in Section~\ref{sec:conclusion}.

\section{IDENTIFICATION OF THE VOLANS-CARINA ASSOCIATION}\label{sec:ident}

\cite{2017AJ....153..257O} identified a list of 8 co-moving \emph{Gaia}~DR1 stars in Volans-Carina, which they \replaced{refer to}{designated} as group~30. \cite{2018arXiv180409058F} verified that the stars from group~30 filled out a logical sequence in a color-magnitude diagram, confirmed this association as newly identified within 100\,\added{p}c, and found\added{ one} of its members \added{(HIP~47017 or HD~83359) }to be detected in ROSAT. In Section~\ref{sec:census} we will use the BANYAN~$\Sigma$ \added{\citep{2018ApJ...856...23G}} Bayesian classification algorithm to uncover new members with \emph{Gaia}~DR2, 2MASS \citep{2006AJ....131.1163S} and AllWISE \citep{2010AJ....140.1868W,2014ApJ...783..122K}. Before building a kinematic model of Volans-Carina and including it in BANYAN~$\Sigma$, we searched \emph{Gaia}~DR2 for any obvious additional members will full kinematics.

We cross-matched the \cite{2017AJ....153..257O} list of group~30 stars with \emph{Gaia}~DR2 and calculated the average Galactic positions $XYZ$ and space velocities $UVW$ of the 5 members that have a radial velocity measurement provided by \emph{Gaia}~DR2\footnote{We weighted individual measurements with their squared inverse errors}. We find average values of $XYZ = (22,-80,-15)$\,pc and $UVW = (-16.0,-27.9,-0.7)$\,\kms. The $XYZUVW$ positions of all \emph{Gaia}~DR2 entries within 125\,pc of the Sun with radial velocity measurements were calculated similarly, which allowed us to identify 13 new objects within 4\,\kms\ (in $UVW$) and 15\,pc (in $XYZ$) that we included in our list of members. This search also recovered the 8 members of \cite{2017AJ....153..257O} in addition to the 13 new objects. We also recovered two A/F-type stars (HD~83523 and HD~83946) that were previously identified by \cite{2011ApJS..192....2S} as co-moving with the \cite{2017AJ....153..257O}\added{ group~30 member HD~83359.}\deleted{ star HD~83359, which is a member of group~30.}

We recovered one additional member of Volans-Carina by doing a similar box search in \emph{Gaia}~DR2 that did not require a radial velocity measurement directly in the \emph{Gaia}~DR2 catalog. To do so, we calculated the average sky position ($\alpha = 145.9$\textdegree, $\delta = -66.1$\textdegree), proper motion ($\mu_\alpha\cos\delta = -37.5$\,\masyr, $\mu_\delta = 43.6$\,\masyr), radial velocity ($22.3$\,\kms) and parallax ($11.8$\,mas) of the members by weighting individual measurements with the squared inverse of their error bars. We then identified any object in the 125\,pc \emph{Gaia}~DR2 sample within 10 degrees, 10\,\masyr\ in both directions of proper motion, and 3\,mas in parallax. The 160 resulting objects were cross-matched with SIMBAD \citep{2000AAS..143...23O} to verify whether they have radial velocity measurements in the literature. Only 3 objects (TYC~9205--1922--1, c~Car and HD~83523) had radial velocity measurements with a precision better than 10\,\kms, and of those three only the B8 star c~Car had a radial velocity ($25.0 \pm 4.1$\,\kms; \citealt{2006AstL...32..759G}) consistent with that of the average Volans-Carina members ($22.3$\,\kms). We therefore added c~Car to the list of members, which are compiled in Table~\ref{tab:orig}. This table does not include the original \cite{2017AJ....153..257O} members HD~83948, HD~82406 and HD~83359, because they do not yet have a radial velocity measurement.

\section{A KINEMATIC MODEL OF VOLANS-CARINA}\label{sec:model}

In this section we build a kinematic model (in $XYZ$ Galactic positions and $UVW$ space velocities) of the Volans-Carina members to include it in the BANYAN~$\Sigma$ Bayesian classification algorithm \citep{2018ApJ...856...23G}. BANYAN~$\Sigma$ uses the sky position, proper motion and optionally radial velocity and/or distance of a star to compare it with the kinematic models of 27 young associations within 150\,pc of the Sun, excluding Volans-Carina, and a model of the field stars within 300\,pc. The kinematic models are composed of one multivariate Gaussian in $XYZUVW$ space for each group, or a mixture of 10 multivariate Gaussians for the field. We used the method described in Section~5 of \cite{2018ApJ...856...23G} to fit a single multivariate Gaussian to the $XYZUVW$ distribution of Volans-Carina members listed in Table~\ref{tab:xyzuvw}.\added{ The BANYAN~$\Sigma$ tool itself is not used to build the kinematic models, but instead will rely on the model to calculate membership probabilities.}

The resulting central position for the Volans-Carina model in $XYZUVW$ space is:
\begin{align}
	\bar x_0 &=  \begin{bmatrix}
    21.6 & -80.3 & -14.3 & -16.11 & -28.13 & -0.85
    \end{bmatrix},\notag
\end{align}
\noindent in units of pc and \kms, and its covariance matrix (in the same units) is:
\begin{align}
    \bar{\bar\Sigma} &= \begin{bmatrix}
    	15 & 2.6 & 1.6 & 0.80 & -0.73 & -0.56 \\
		2.6 & 25 & 2.6 & 4.4 & 2.2 & 0.77 \\
        1.6 & 2.6 & 20 & 0.93 & 1.1 & -1.5 \\
        0.80 & 4.4 & 0.94 & 2.0 & 0.48 & 0.39\\
        -0.73 & 2.2 & 1.1 & 0.48 & 2.0 & -0.16\\
        -0.56 & 0.77 & -1.5 & 0.39 & -0.16 & 1.0
    \end{bmatrix}.\notag
\end{align}

The 1$\sigma$ contours of the Gaussian model projected on the $XYZUVW$ axes are the square root of the diagonal elements of $\bar{\bar\Sigma}$, and the off-diagonal elements of $\bar{\bar\Sigma}$ represent the covariances in the distribution of members in 6-dimensional space.

The construction of BANYAN~$\Sigma$ kinematic models also includes a Monte Carlo calculation (see Section~7 of \citealt{2018ApJ...856...23G}) where $10^7$ random synthetic stars are drawn and where a Bayesian prior is adjusted such that 50\%, 68\%, 82\% or 90\% of the true members are recovered with a Bayesian probability $P > 90$\% when the input measurements include (1) sky position and proper motion only, (2) sky position, proper motion and radial velocity, (3) sky position, proper motion and parallax, or (4) sky position, proper motion, radial velocity and parallax, respectively. This is done to make BANYAN~$\Sigma$ easier to use with a single probability threshold that generates a fixed recovery rate of true members across all associations. We found the following values for the natural logarithm of these Bayesian priors for Volans-Carina (referred to as $\ln\alpha_k$ in \citealt{2018ApJ...856...23G}) in the order corresponding to the scenarios of input data described above:
\begin{align}
	\ln\alpha_k &=  \begin{bmatrix}
    -17.2 & -18.7 & -21.1 & -22.0
    \end{bmatrix}.\notag
\end{align}

Other relevant characteristics of Volans-Carina are listed in Table~\ref{tab:vca}, and its new model was included in version 1.2 of BANYAN~$\Sigma$ in both the IDL and Python versions of the algorithm\footnote{An IDL version is available at \url{https://github.com/jgagneastro/banyan_sigma_idl} and a Python version at \url{https://github.com/jgagneastro/banyan_sigma} \citep{zenodobanyansigmapython,zenodobanyansigmaidl}, and a web tool is available at \url{www.exoplanetes.umontreal.ca/banyan/banyansigma.php}.}, where the Volans-Carina association is abbreviated as VCA.

\begin{deluxetable}{lc}
\renewcommand\arraystretch{0.9}
\tabletypesize{\small}
\tablecaption{Fundamental and Average Properties of the Volans-Carina association \label{tab:vca}}
\tablehead{\colhead{Property} & \colhead{Value}}
\startdata
Age (Myr) & \age\\
Members & \likelymembers\\
Candidate members & \candidatemembers\\
Spatial size\tablenotemark{a} (pc) & 4.4\\
Kinematic size (\kms) & 1.2\\
$\left<{\rm R.A.}\right>$ & 09:37:00\,$\pm$\,2.7\,\textdegree\\
$\left<{\rm Decl.}\right>$  & --65:18:00\,$\pm$\,2.8\,\textdegree\\
$\left<\mu_\alpha\cos\delta\right>$ (\masyr) & $-36 \pm 7$\\
$\left<\mu_\delta\right>$ (\masyr) & $42 \pm 6$\\
$\left<{\rm RV}\right>$ (\kms) & $23 \pm 2$\\
$\left<{\rm Trig. dist.}\right>$ (pc) & $86 \pm 5$\\
$\left<{\rm Parallax}\right>$ (mas) & $11.6 \pm 0.6$\\
$\left<X\right>$ (pc) & $20.5 \pm 3.0$\\
$\left<Y\right>$ (pc) & $-81.6 \pm 4.8$\\ 
$\left<Z\right>$ (pc) & $-4.6 \pm 8.5$\\
$\left<U\right>$ (\kms) & $-16.14 \pm 0.88$\\
$\left<V\right>$ (\kms) & $-28.4 \pm 1.1$\\
$\left<W\right>$ (\kms) & $-1.1 \pm 1.2$\\
\enddata
\tablenotetext{a}{The spatial and kinematic sizes are defined as the radius of a sphere that would have the same volume as the 1$\sigma$ contour of the $XYZ$ or $UVW$ projections of the 6-dimensional BANYAN~$\Sigma$ kinematic model.}
\tablecomments{All average values $\left<X\right>$ are obtained with a weighted average of the individual measurements of the members listed in Table~\ref{tab:orig}, and their error bars are obtained with an un-biased weighted standard deviation. In both cases, the weights are set to the squared of the individual inverse error bars. The error bars are representative of the intrinsic scatter, not an error on the average. See \cite{2018ApJ...856...23G} for more details.}
\end{deluxetable}

\section{COMPLETING THE CENSUS OF CANDIDATE MEMBERS}\label{sec:census}

In this Section, we use the BANYAN~$\Sigma$ Bayesian classifier algorithm to identify additional candidate members of the Volans-Carina association. We perform two separate searches: the first one is based on the full \emph{Gaia}~DR2 nearest 125\,pc sample, and is appropriate to identify the previously missing stellar members of Volans-Carina, especially among the \emph{Gaia}~DR2 entries that do not have a radial velocity measurement. The second search is based on a cross-match of the 2MASS \citep{2006AJ....131.1163S} and AllWISE \citep{2010AJ....140.1868W,2014ApJ...783..122K} surveys, and is appropriate to identify the substellar members too faint to be detected in \emph{Gaia}~DR2.

\begin{figure}
	\centering
	\includegraphics[width=0.465\textwidth]{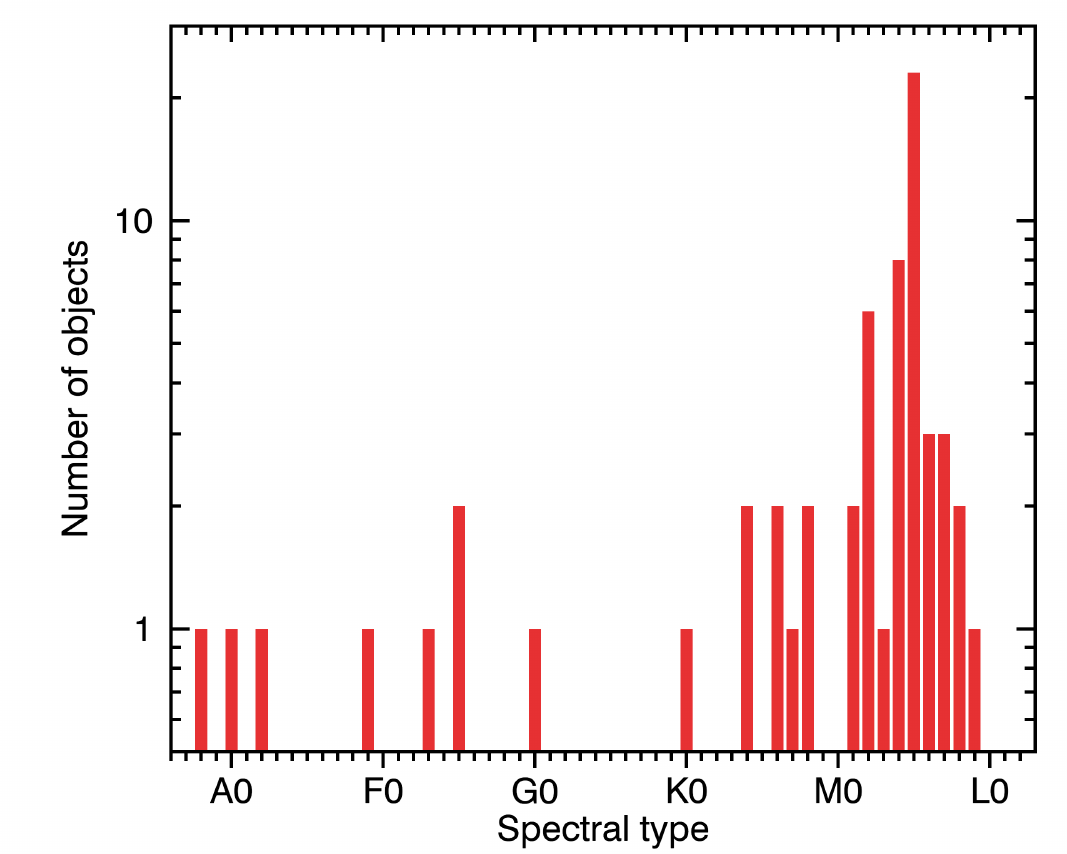}
	\caption{Spectral types histogram of Volans-Carina members and candidates. The census of members is dominated by M-type dwarfs, which is expected for a typical initial mass function. See Section~\ref{sec:census} for more detail.}
	\label{fig:spt}
\end{figure}

\subsection{A \emph{Gaia}~DR2 Search for Stellar Members}\label{sec:dr2census}

The 1\,427\,111 \emph{Gaia}~DR2 entries within 125\,pc\added{\footnote{including those without radial velocity measurements}} were analyzed with the BANYAN~$\Sigma$ Bayesian classification algorithm to identify additional candidate members of the Volans-Carina association. The 58 entries with Volans-Carina Bayesian membership probabilities above 90\% and a predicted $UVW$ position within 5\,\kms\ of the locus of Volans-Carina members were selected.

The spectral types of the new candidate members recovered here were estimated based on their absolute \emph{Gaia}~DR2 $G$-band magnitude with the relations of \cite{2018ApJ...862..138G}, and they are reported in Table~\ref{tab:cms} with the list of candidate members.

\subsection{A 2MASS--AllWISE Search for Substellar Members}

We selected all 2MASS and AllWISE entries located between \replaced{R.A. 127--155.5\textdegree and Decl. $-$72.3..$-$57\textdegree}{127$^{\circ}$ $<$ $\alpha$ $<$
-155$^{\circ}$.5 and -72$^{\circ}$.3 $<$ $\delta$ $<$
-57$^{\circ}$}, and performed a cross-match where each 2MASS entry was paired with its nearest AllWISE neighbor. A proper motion was then calculated for each of these 5\,136\,857 matches based on the 2MASS and AllWISE astrometries, with the method described by \cite{2015ApJ...798...73G}. The sky positions and proper motions of all matches were analyzed with the BANYAN~$\Sigma$ classification algorithm, and the 2\,927 entries with a Bayesian probability above 90\% and located at minimum distance of less than 5\,\kms\ from the locus of Volans-Carina members in $UVW$ space were selected for further consideration.

All resulting 2MASS entries were cross-matched with \emph{Gaia}~DR2 by projecting back the \emph{Gaia} entries at epoch 2000 (using the sky positions and proper motions of \emph{Gaia}~DR2) with a cross-match radius of 2$''$. All 2\,884 matches in \emph{Gaia}~DR2 with parallaxes were rejected because they were already investigated in Section~\ref{sec:dr2census}. A visual inspection of near-infrared color-magnitude diagrams at the average distance of Volans-Carina (86\,pc) for the 43 remaining candidate members allowed us to reject 29 of them that have photometric properties inconsistent with substellar objects of either young or field ages (i.e., those with $W1-W2 < 0.1$ and $M_{W1} >8.5$; or those with $J-K_S <0.8$ were rejected; see \citealp{2011ApJS..197...19K,2012ApJS..201...19D,2016ApJS..225...10F}).

The Digitized Sky Survey, 2MASS and \emph{WISE} images of the remaining sources were inspected, allowing us to reject an additional 11/14 that were bright at visible wavelengths, extended sources or clearly contaminated by neighbors.\added{ Such rejection criteria are typical of brown dwarf surveys (e.g., \citealp{2011ApJS..197...19K,2016ApJ...830..144R}), and we rely on them because our 2MASS--AllWISE candidates not in \emph{Gaia}~DR2 have colors typical of substellar objects. The first two criteria are especially useful to distinguish nearby substellar objects from distant extragalactic contaminants, which can otherwise have similar near-infrared colors.} The remaining three objects (2MASS~J09254243--6408524, 2MASS~J09285886--6541371 and 2MASS~J09374003--6226201) are listed in Table~\ref{tab:cms} and have respective spectral type estimates M7, M9 and M6 based on their absolute $K_S$-band magnitudes at the average distance of Volans-Carina compared with the spectral type--absolute magnitude relations of \cite{2016ApJS..225...10F}. A spectroscopic follow-up will be needed to confirm their substellar nature, and measure their radial velocities and spectroscopic signs of youth to confirm their membership in the Volans-Carina association.

\begin{figure}
	\centering
	\includegraphics[width=0.465\textwidth]{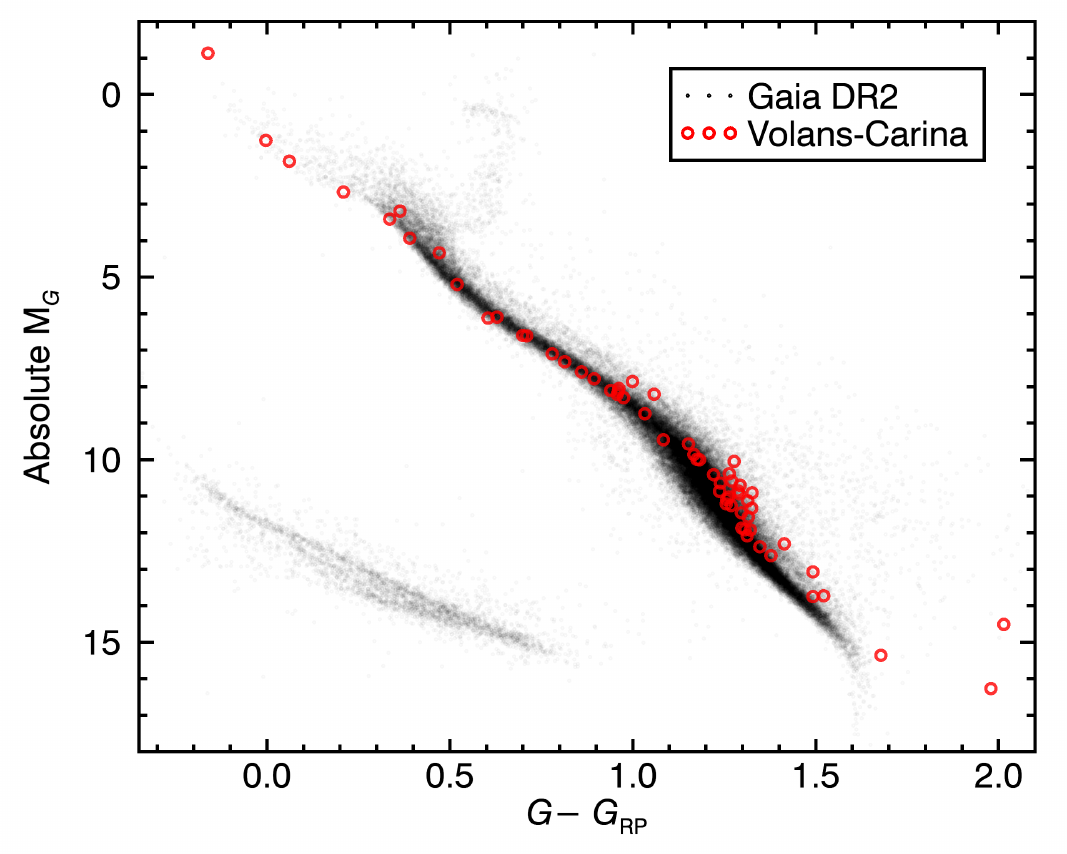}
	\caption{\emph{Gaia} absolute $G$ versus $G-G_{\rm RP}$ magnitude for the 100\,pc sample of high-quality parallaxes in \emph{Gaia}~DR2 (black dots) compared to the proposed members and candidate members of the Volans-Carina association (red empty circles). See Section~\ref{sec:census} for more detail.}
	\label{fig:volans_cmd}
\end{figure}

\section{DISCUSSION}\label{sec:discussion}

In this Section, we investigate the fundamental properties of the Volans-Carina members, including their\deleted{ metallicity (Section~\ref{sec:feh}),} age (Section~\ref{sec:age}), chromospheric activity (Section~\ref{sec:activ}) and present-day mass function (Section~\ref{sec:imf}). We also discuss a white dwarf interloper that has a cooling age that is too old for it to be a member of Volans-Carina (Section~\ref{sec:wd}), and the similarity between the age and kinematics of Volans-Carina and the Platais~8 association (Section~\ref{sec:relat}).

\startlongtable
 \begin{deluxetable*}{llcccccc}
\tablecolumns{8}
\tabletypesize{\scriptsize}
\tablecaption{Galactic positions and space velocities of Volans-Carina members.\label{tab:xyzuvw}}
 \tablehead{\colhead{VCA} & \colhead{Name} & \colhead{$X$} & \colhead{$Y$} & \colhead{$Z$} & \colhead{$U$} & \colhead{$V$} & \colhead{$W$}
\\
 \colhead{ID} & \colhead{} & \colhead{(pc)} & \colhead{(pc)} & \colhead{(pc)} & \colhead{(\kms)} & \colhead{(\kms)} & \colhead{(\kms)}
 }
\startdata
1 & c Car & $12.29 \pm 0.39$ & $-91.5 \pm 2.9$ & $-16.26 \pm 0.52$ & $-19.44 \pm 0.83$ & $-27.8 \pm 2.4$ & $-2.22 \pm 0.51$\\
2 & HD 80563 & $17.976 \pm 0.045$ & $-88.30 \pm 0.22$ & $-15.578 \pm 0.039$ & $-16.66 \pm 0.10$ & $-27.89 \pm 0.41$ & $-0.582 \pm 0.075$\\
3 & HD 83946 & $22.109 \pm 0.054$ & $-86.57 \pm 0.21$ & $-14.680 \pm 0.036$ & $-16.68 \pm 0.25$ & $-27.82 \pm 0.97$ & $-0.61 \pm 0.17$\\
4 & HD 309681 & $19.18 \pm 0.20$ & $-83.18 \pm 0.87$ & $-15.18 \pm 0.16$ & $-14.35 \pm 0.26$ & $-30.94 \pm 0.45$ & $0.64 \pm 0.15$\\
5 & CD-67 852 & $26.047 \pm 0.049$ & $-76.49 \pm 0.15$ & $-13.489 \pm 0.026$ & $-16.18 \pm 0.12$ & $-28.31 \pm 0.31$ & $-0.787 \pm 0.057$\\
6 & TYC 8933-327-1 & $10.636 \pm 0.084$ & $-73.26 \pm 0.58$ & $-19.12 \pm 0.15$ & $-13.19 \pm 0.30$ & $-26.1 \pm 1.8$ & $0.70 \pm 0.47$\\
7 & TYC 8953-1289-1 & $22.227 \pm 0.046$ & $-79.81 \pm 0.17$ & $-15.547 \pm 0.033$ & $-16.35 \pm 0.15$ & $-27.87 \pm 0.49$ & $-0.797 \pm 0.098$\\
8 & TYC 8950-1447-1 & $19.105 \pm 0.034$ & $-74.01 \pm 0.13$ & $-10.687 \pm 0.019$ & $-16.28 \pm 0.11$ & $-28.32 \pm 0.39$ & $-0.903 \pm 0.059$\\
9 & TYC 9210-1818-1 & $27.177 \pm 0.052$ & $-81.10 \pm 0.15$ & $-14.712 \pm 0.028$ & $-14.43 \pm 0.19$ & $-27.13 \pm 0.54$ & $-0.39 \pm 0.10$\\
10 & 2MASS J09201987-6607418 & $20.329 \pm 0.043$ & $-83.18 \pm 0.17$ & $-17.396 \pm 0.036$ & $-16.27 \pm 0.15$ & $-28.53 \pm 0.60$ & $-0.82 \pm 0.13$\\
11 & 2MASS J09345577-6500076 & $20.743 \pm 0.060$ & $-83.07 \pm 0.24$ & $-14.516 \pm 0.042$ & $-15.82 \pm 0.32$ & $-29.3 \pm 1.3$ & $-2.07 \pm 0.22$\\
12 & 2MASS J09330015-6251207 & $18.794 \pm 0.039$ & $-85.63 \pm 0.18$ & $-12.635 \pm 0.026$ & $-16.77 \pm 0.41$ & $-26.5 \pm 1.8$ & $-0.59 \pm 0.27$\\
13 & 2MASS J09263145-6236236 & $18.01 \pm 0.18$ & $-87.23 \pm 0.89$ & $-13.37 \pm 0.14$ & $-18.44 \pm 0.51$ & $-30.7 \pm 2.1$ & $-0.60 \pm 0.34$\\
14 & 2MASS J08591364-6919442 & $21.58 \pm 0.15$ & $-81.32 \pm 0.55$ & $-22.69 \pm 0.15$ & $-16.59 \pm 0.33$ & $-27.5 \pm 1.1$ & $-1.44 \pm 0.31$\\
15 & 2MASS J10223868-5847599 & $20.977 \pm 0.051$ & $-80.10 \pm 0.19$ & $-1.8878 \pm 0.0046$ & $-14.44 \pm 0.81$ & $-26.6 \pm 3.1$ & $-3.772 \pm 0.075$\\
16 & 2MASS J08591527-6918426 & $21.609 \pm 0.039$ & $-81.52 \pm 0.15$ & $-22.726 \pm 0.041$ & $-16.69 \pm 0.33$ & $-29.2 \pm 1.2$ & $-2.14 \pm 0.35$\\
17 & 2MASS J09311626-6414270 & $19.849 \pm 0.058$ & $-84.62 \pm 0.25$ & $-14.293 \pm 0.042$ & $-16.08 \pm 0.70$ & $-30.4 \pm 3.0$ & $-0.94 \pm 0.51$\\
18 & 2MASS J09174971-6628140 & $21.983 \pm 0.075$ & $-89.42 \pm 0.31$ & $-19.409 \pm 0.067$ & $-16.14 \pm 0.55$ & $-29.9 \pm 2.2$ & $-1.25 \pm 0.48$\\
19 & 2MASS J09203649-6309598 & $18.150 \pm 0.036$ & $-88.54 \pm 0.17$ & $-14.976 \pm 0.029$ & $-16.59 \pm 0.92$ & $-27.9 \pm 4.5$ & $-0.70 \pm 0.76$\\
\enddata
\end{deluxetable*}

\subsection{The Age of the Volans-Carina Association}\label{sec:age}

We used a Bayesian method to determine a probability density for the age of Volans-Carina, based on the fiducial MIST solar-metallicity model isochrones of \cite{2016ApJ...823..102C} that include stellar rotation ($v = 0.4 v_{\rm crit}$) and were generated with the revised \emph{Gaia}~DR2 photometric zero points of \cite{2018arXiv180409368E}\footnote{The corrected models are available at \url{http://waps.cfa.harvard.edu/MIST/model_grids.html} since 2018 April 27, 3:30PM EST.}. It is well known that the model-derived bolometric corrections as a function of effective temperature have systematic biases that depend on spectral types and surface gravity (i.e., mass and age). Therefore, relying on the models without calibration would generate biased ages (e.g., see \citealt{2015MNRAS.454..593B}). Additional potential causes of systematic errors include the effect of magnetic fields that inflate the radius of young low-mass stars (e.g., see \citealp{2014ApJ...792...37M,2016AA...593A..99F}) and the effect of increased chromospheric activity on the colors of K- and possibly later-type dwarfs \citep{2003AJ....126..833S}.

\begin{figure*}
	\centering
	\includegraphics[width=0.965\textwidth]{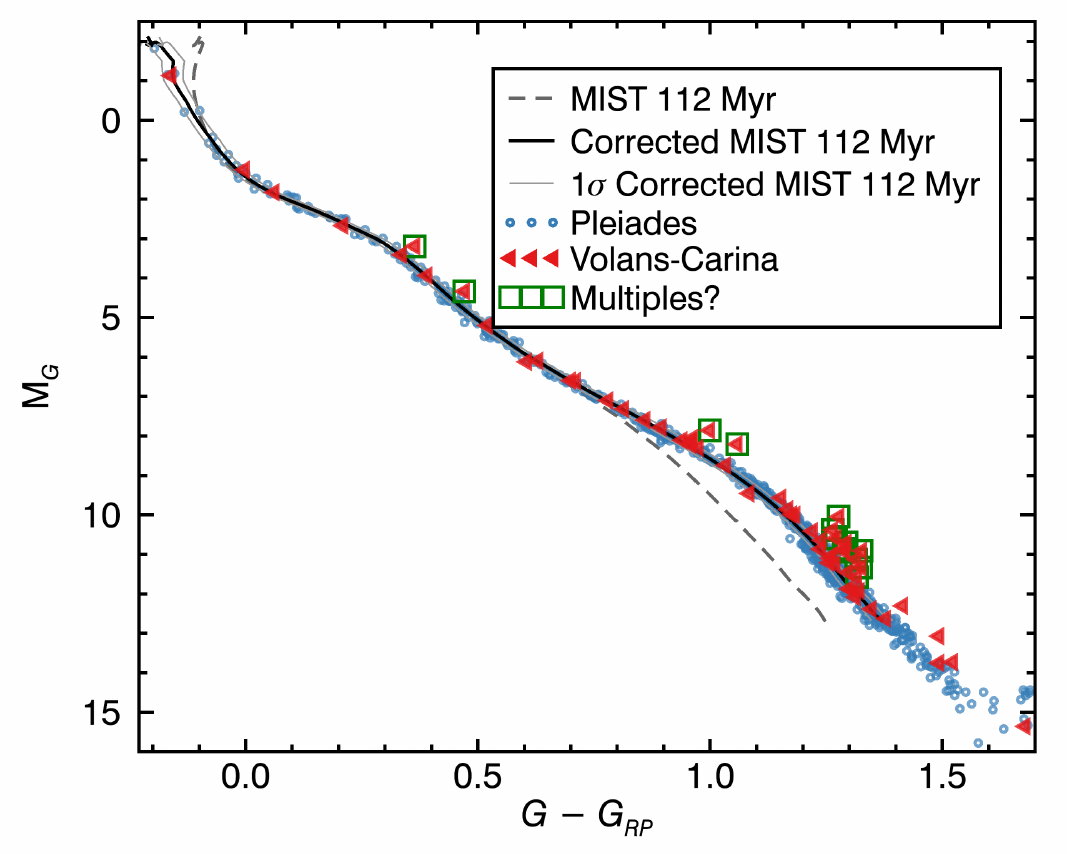}
	\caption{\emph{Gaia} absolute $G$ versus $G - G_{\rm RP}$ magnitude for members of Volans-Carina (blue leftward triangles) and the Pleiades (red circles) compared to the $112$\,Myr solar-metallicity MIST isochrone (dashed gray line), and the empirically corrected isochrone (thick black line). The 1$\sigma$ uncertainties in the corrected isochrone resulting from the intrinsic spread in the Pleiades members are displayed as thin gray lines, and the potential unresolved multiples are marked with green squares. Members of the Pleiades were de-reddened with the relation of \cite{2008AJ....136.1388T}, and the over-luminous Pleiades members that likely correspond to multiple systems were removed.\replaced{ Members of Volans-Carina were de-reddened with $E(B-V) = 0.008$\,mag with the exception of c~Car ($E(B-V)= 0.023$\,mag).}{ Members of Volans-Carina have negligible reddening.} The sequence formed by the members of Volans-Carina is similar to that of the Pleiades members, indicating that the two associations are likely coeval. See Section~\ref{sec:age} for more detail.}
	\label{fig:volans_ple_isoc}
\end{figure*}

One way to circumvent these systematic biases is to use associations with a well-calibrated age to apply color corrections on the model isochrones, however in general this still requires assumptions on how these corrections depend on age. Fortunately, the members of Volans-Carina form a sequence that is remarkably close to that of the Pleiades ($112 \pm 5$\,Myr; \citealt{2015ApJ...813..108D}) in color-magnitude diagrams (e.g., see Figure~\ref{fig:volans_ple_isoc}), which indicates that the two associations have a similar age. We can therefore derive an absolute magnitude-dependent color correction of the model isochrones based on the Pleiades members that will be valid for ages in the vicinity $\simeq$\,112\,Myr. Model isochrones at very different ages will yield small probabilities of matching all members of Volans-Carina, and therefore applying an age-dependence on the Pleiades-derived color correction would only have a small effect on our age determination.

Members of the Pleiades were de-reddened with the relation of \cite{2008AJ....136.1388T}; i.e., $E(B-V) = 0.057 \pm 0.008$\,mag for R.A.\,$<$\,56.819\textdegree, and $E(B-V) = 0.034 \pm 0.011$\,mag otherwise.\deleted{ A cross-match of our list of Volans-Carina members with \cite{2006ApJ...638.1004A} allowed us to obtain reddening estimates for 5 objects, with a median $E(B-V) = 0.008$\,mag, and median absolute deviation of $E(B-V) = 0.001$\,mag: we therefore de-reddened the members of Volans-Carina using this value. In the case of c~Car, which is spatially located in the outskirts of Volans-Carina, we used the revised Q-method of \cite{2013ApJS..208....9P} with the mean photometry of \cite{1987AAS...71..413M}\footnote{$V = 3.835 \pm 0.005$, $B-V =
-0.103 \pm 0.005$ and $U-B = -0.440 \pm 0.006$; see \url{http://vizier.u-strasbg.fr/viz-bin/VizieR-5?-ref=VIZ5b0eda1fb49b&-out.add=.&-source=II/168/ubvmeans&recno=22342}.} to estimate a reddening of $E(B-V)= 0.023$\,mag.}\added{ The reddening map of \cite{2017AA...606A..65C} and the distance and position of VCA indicates a reddening of $E(B-V) = 0.009 \pm 0.016$\,mag. We also identified 5 stars\footnote{HD~77370 (26\,pc;
$E(b-y) = -0.004 \pm 0.009$), HD~75171 (59\,pc; $E(b-y) =
-0.006 \pm 0.006$), HD~79629 (83\,pc; $E(b-y) =
0.005 \pm 0.007$), HD~80563 (VCA~2; 91\,pc; $E(b-y) =
-0.003 \pm 0.012$), HD~82244 (112\,pc; $E(b-y) =
0.008 \pm 0.014$).} in the reddening catalog of \cite{2011ApJ...734....8R} in the direction of VCA at distances 26--112\,pc. All of them have a negligible reddening within the error bars, consistent with the reddening map of \cite{2017AA...606A..65C}. We therefore consider that the members of VCA are subject to a negligible amount of reddening from interstellar extinction.}

\added{The Q-method de-reddening calibration for dwarf stars from \cite{2013ApJS..208....9P} is inappropriate for c~Car as the star is a supergiant
(B8\,II with Mn metallic lines; \citealt{1994AJ....107.1556G}),
and has been classified by other authors since the 1950s as class II
or III (B8\,III; \citealt{1977AAS...30...71C}; B8\,II; \citealt{1957MNRAS.117..449D}).
\cite{1979PASP...91..299G} shows that the there is
considerable spread in intrinsic ($U-B$) and ($B-V$) colors between
supergiants and giants, however the Johnson colors or c~Car ($U-B =
-0.440 \pm 0.006$, $B-V = -0.103 \pm 0.005$; \citealt{1994cmud.book.....M}\footnote{Vizier catalog II/168/ubvmeans}) appear to be consistent with a negligibly reddened
star with type between B7 and B8 and luminosity class II and III.
\cite{1980AAS...40..199P} dereddened the $uvby$
photometry and estimated color excess of $E(b-y) = 0.009$, consistent
with $A_v = 0.04$ following the calibration of \cite{1971AJ.....76.1041G}.}

\begin{figure}
	\centering
	\includegraphics[width=0.465\textwidth]{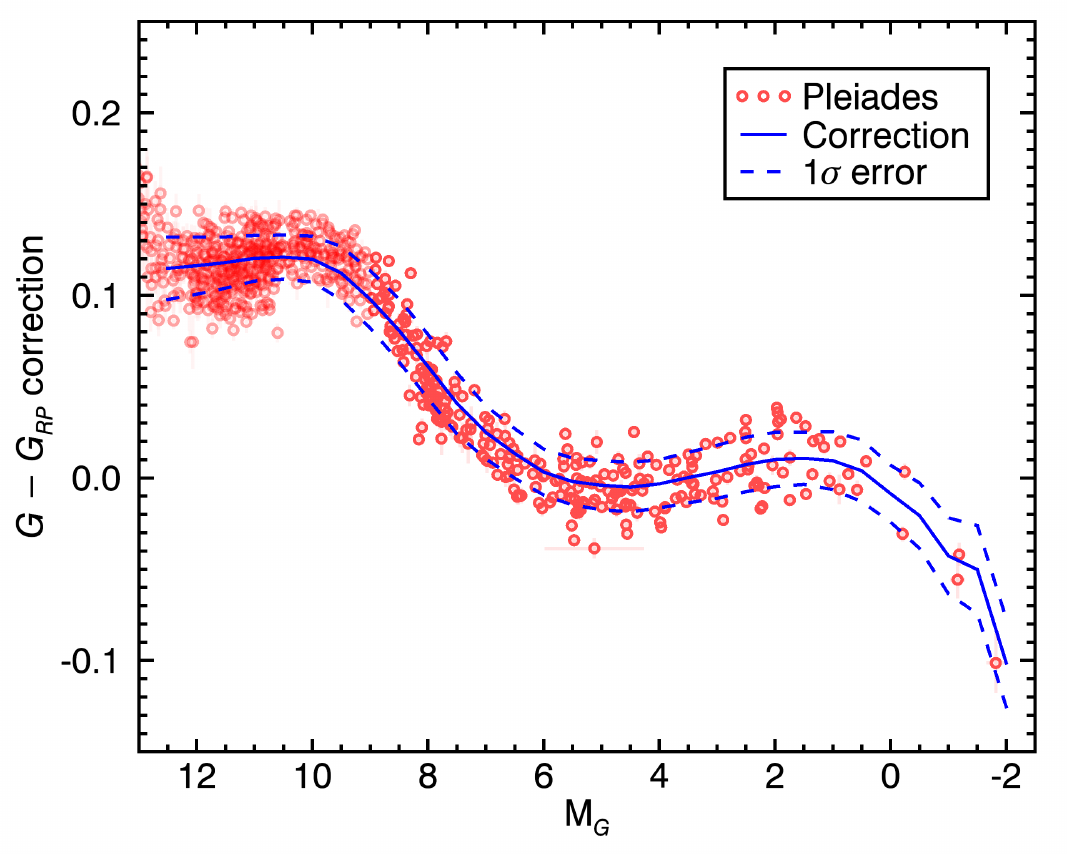}
	\caption{Empirical color correction in $G - G_{\rm RP}$ color versus $G$-band absolute magnitude based on a comparison of Pleiades members (red circles) with the $112$\,Myr solar-metallicity MIST isochrone. The average and 1$\sigma$ errors on the color correction are displayed as full and dashed blue lines, respectively. The data behind this figure are available in the online version of this paper. See Section~\ref{sec:age} for more detail.}
	\label{fig:ple_corr}
\end{figure}

In Figure~\ref{fig:ple_corr}, we show deviations in the $G - G_{\rm RP}$ colors of Pleiades members compared to the $112$\,Myr MIST isochrone, as a function of absolute $G$-band magnitude. We used a sliding weighted average to derive a color correction in steps of 0.5\,mag and a running box width of 1\,mag. We used weights proportional to the squared inverse of the error bars on the colors of individual members. A 1$\sigma$ error bar was also calculated at each point of the sequence by using a sliding un-biased weighted standard deviation\footnote{See \url{https://www.gnu.org/software/gsl/manual/html_node/Weighted-Samples.html}} with the same weights and box width. The average color correction and its error bars were then smoothed with a sliding average using a box width of 3 array elements. The resulting color correction is displayed in Figure~\ref{fig:ple_corr}, and is only valid for the \emph{Gaia}~DR2 absolute $G$-band versus $G - G_{\rm RP}$ color-magnitude diagram, for stars with colors in the range $-$0.18..1.37\,mag and absolute magnitudes in the range $-$2..12.6\,mag.

\startlongtable
 \begin{deluxetable*}{llcccccc}
\tablecolumns{8}
\tablecaption{Candidate Members of Volans-Carina.\label{tab:cms}}
 \tablehead{\colhead{2MASS} & \colhead{Spec.} & \colhead{R.A.\tablenotemark{b}} & \colhead{Decl.\tablenotemark{b}} & \colhead{$\mu_\alpha\cos\delta$} & \colhead{$\mu_\delta$} & \colhead{Parallax} & \colhead{Mem. Prob.\tablenotemark{c}}
 \\
 \colhead{ID} & \colhead{Type\tablenotemark{a}} & \colhead{(hh:mm:ss.sss)} & \colhead{(dd:mm:ss.ss)} & \colhead{(\masyr)} & \colhead{(\masyr)} & \colhead{(mas)} & \colhead{(\%)}
}
\startdata
08383351-6716368 & (M5) & 08:38:33.458 & -67:16:35.95 & $-18.69 \pm 0.16$ & $53.86 \pm 0.18$ & $11.66 \pm 0.09$ & 99.3\\
08485563-6113261 & (M5) & 08:48:55.579 & -61:13:25.38 & $-22.47 \pm 0.16$ & $45.64 \pm 0.17$ & $11.43 \pm 0.08$ & 97.6\\
08511579-7140152 & (K8) & 08:51:15.710 & -71:40:14.37 & $-24.49 \pm 0.04$ & $60.56 \pm 0.04$ & $12.61 \pm 0.02$ & 94.2\\
08524066-6552278\tablenotemark{d} & (M4) & 08:52:40.594 & -65:52:27.06 & $-21.29 \pm 0.11$ & $48.32 \pm 0.11$ & $11.23 \pm 0.06$ & 99.5\\
08524155-6401229 & (M6) & 08:52:41.501 & -64:01:22.14 & $-24.97 \pm 0.19$ & $49.68 \pm 0.23$ & $12.03 \pm 0.09$ & 99.4\\
08544569-7055078 & (M5) & 08:54:45.629 & -70:55:07.08 & $-25.42 \pm 0.09$ & $59.88 \pm 0.10$ & $12.31 \pm 0.06$ & 96.9\\
08555655-6146057 & (M5) & 08:55:56.496 & -61:46:05.11 & $-23.05 \pm 0.14$ & $43.65 \pm 0.14$ & $10.75 \pm 0.07$ & 97.0\\
09040687-6325337 & (M5) & 09:04:06.799 & -63:25:32.92 & $-29.42 \pm 0.10$ & $50.47 \pm 0.08$ & $12.81 \pm 0.05$ & 98.4\\
09090086-6826022\tablenotemark{d} & (M5) & 09:09:00.787 & -68:26:01.52 & $-24.6 \pm 1.2$ & $50.7 \pm 1.2$ & $13.26 \pm 0.75$ & 98.9\\
09111442-6631389 & (M4) & 09:11:14.359 & -66:31:38.18 & $-29.16 \pm 0.09$ & $47.69 \pm 0.09$ & $11.59 \pm 0.04$ & 99.9\\
09133635-6522114 & (M5) & 09:13:36.286 & -65:22:10.77 & $-28.70 \pm 0.09$ & $46.31 \pm 0.09$ & $11.36 \pm 0.05$ & 99.9\\
09181573-6310540 & (M5) & 09:18:15.672 & -63:10:53.32 & $-30.81 \pm 0.10$ & $43.90 \pm 0.10$ & $11.60 \pm 0.06$ & 99.8\\
09191188-6640123\tablenotemark{d} & (M6) & 09:19:11.796 & -66:40:11.47 & $-35.45 \pm 0.27$ & $53.86 \pm 0.27$ & $13.36 \pm 0.13$ & 97.3\\
09202620-6329547\tablenotemark{d} & (M7) & 09:20:26.126 & -63:29:54.15 & $-29.74 \pm 0.42$ & $42.14 \pm 0.46$ & $10.99 \pm 0.21$ & 99.6\\
09221992-6756519 & (M4) & 09:22:19.841 & -67:56:51.15 & $-32.11 \pm 0.07$ & $48.84 \pm 0.07$ & $11.70 \pm 0.04$ & 99.9\\
09223111-6206070\tablenotemark{d} & (M5) & 09:22:31.037 & -62:06:06.57 & $-31.41 \pm 0.19$ & $40.42 \pm 0.17$ & $11.16 \pm 0.09$ & 99.6\\
09234200-6556512 & (M4) & 09:23:41.916 & -65:56:50.53 & $-32.52 \pm 0.06$ & $46.24 \pm 0.06$ & $11.66 \pm 0.03$ & 99.9\\
09244301-6254468 & (M5) & 09:24:42.943 & -62:54:46.14 & $-30.83 \pm 0.13$ & $42.19 \pm 0.13$ & $11.08 \pm 0.07$ & 99.7\\
09244337-6856223 & (M5) & 09:24:43.270 & -68:56:21.48 & $-32.35 \pm 0.11$ & $50.46 \pm 0.11$ & $11.53 \pm 0.07$ & 99.8\\
09244959-6216319 & (M7) & 09:24:49.519 & -62:16:31.28 & $-30.27 \pm 0.35$ & $39.48 \pm 0.29$ & $11.40 \pm 0.17$ & 99.3\\
09254243-6408524 & (M7) & 09:25:42.362 & -64:08:51.77 & $-30.7 \pm 4.9$ & $45.2 \pm 8.1$ & $\cdots$ & 96.8\\
09280826-6553589 & (M5) & 09:28:08.165 & -65:53:58.23 & $-30.59 \pm 0.16$ & $42.26 \pm 0.14$ & $10.55 \pm 0.08$ & 98.7\\
09283051-6642067 & A0V & 09:28:30.456 & -66:42:05.99 & $-32.88 \pm 0.28$ & $46.76 \pm 0.27$ & $11.93 \pm 0.15$ & 99.9\\
09285886-6541371 & (M9) & 09:28:58.791 & -65:41:36.48 & $-32.5 \pm 9.0$ & $45 \pm 16$ & $\cdots$ & 92.8\\
$\cdots$ & (M5) & 09:29:31.111 & -63:45:37.15 & $-30.95 \pm 0.11$ & $43.41 \pm 0.09$ & $11.41 \pm 0.05$ & 99.7\\
09293121-6345391\tablenotemark{f} & (M4) & 09:29:31.147 & -63:45:39.03 & $-34.81 \pm 0.08$ & $41.77 \pm 0.07$ & $11.32 \pm 0.04$ & 99.9\\
09312193-6419239\tablenotemark{d} & (M5) & 09:31:21.845 & -64:19:23.40 & $-31.45 \pm 0.14$ & $41.15 \pm 0.15$ & $10.34 \pm 0.08$ & 97.0\\
09313619-6659363 & (M8) & 09:31:36.103 & -66:59:35.59 & $-37.5 \pm 1.1$ & $48.29 \pm 0.91$ & $11.48 \pm 0.64$ & 99.7\\
09314405-6600538 & (M5) & 09:31:43.946 & -66:00:53.09 & $-38.72 \pm 0.11$ & $52.21 \pm 0.10$ & $13.53 \pm 0.06$ & 97.2\\
09323325-6908439\tablenotemark{e} & (M5) & 09:32:33.149 & -69:08:43.18 & $-34.15 \pm 0.15$ & $48.30 \pm 0.19$ & $11.53 \pm 0.08$ & 99.8\\
09330028-6330381 & (M4) & 09:33:00.187 & -63:30:37.54 & $-34.34 \pm 0.06$ & $41.41 \pm 0.07$ & $11.29 \pm 0.03$ & 99.9\\
09345645-6459579\tablenotemark{d} & F5V & 09:34:56.376 & -64:59:57.38 & $-34.70 \pm 0.06$ & $43.97 \pm 0.05$ & $11.44 \pm 0.03$ & 99.9\\
09360518-6457011 & A2V & 09:36:05.098 & -64:57:00.25 & $-35.08 \pm 0.08$ & $40.57 \pm 0.07$ & $11.39 \pm 0.04$ & 99.8\\
09374003-6226201 & (M6) & 09:37:39.936 & -62:26:19.50 & $-46.6 \pm 5.3$ & $42.3 \pm 7.3$ & $\cdots$ & 93.0\\
09384522-6651326 & A9IV/V & 09:38:45.130 & -66:51:31.99 & $-36.61 \pm 0.05$ & $45.09 \pm 0.05$ & $11.69 \pm 0.03$ & 99.9\\
09410683-6658584\tablenotemark{d} & (M5) & 09:41:06.749 & -66:58:57.79 & $-32.51 \pm 0.12$ & $39.79 \pm 0.11$ & $10.44 \pm 0.07$ & 91.6\\
09430009-6422290\tablenotemark{d} & (M5) & 09:43:00.014 & -64:22:28.53 & $-33.57 \pm 0.12$ & $38.83 \pm 0.13$ & $10.45 \pm 0.07$ & 96.3\\
09451524-6908233 & (M3) & 09:45:15.134 & -69:08:22.66 & $-39.16 \pm 0.04$ & $47.61 \pm 0.05$ & $12.03 \pm 0.02$ & 99.8\\
09522444-6731136\tablenotemark{d} & (M5) & 09:52:24.343 & -67:31:12.96 & $-39.60 \pm 0.13$ & $43.99 \pm 0.14$ & $11.48 \pm 0.08$ & 99.9\\
09575077-6415362 & (M8) & 09:57:50.676 & -64:15:35.51 & $-40.61 \pm 0.90$ & $40.41 \pm 0.92$ & $10.75 \pm 0.44$ & 99.1\\
10014811-6508434 & (M4) & 10:01:48.014 & -65:08:42.82 & $-43.77 \pm 0.08$ & $41.02 \pm 0.08$ & $11.97 \pm 0.05$ & 99.9\\
10024136-6845179 & (M4) & 10:02:41.256 & -68:45:17.33 & $-41.62 \pm 0.05$ & $38.07 \pm 0.05$ & $10.67 \pm 0.03$ & 95.1\\
10082205-6757040 & (M5) & 10:08:21.938 & -67:57:03.45 & $-37.94 \pm 0.10$ & $38.79 \pm 0.10$ & $11.46 \pm 0.06$ & 97.7\\
10101695-6616006\tablenotemark{d} & (M5) & 10:10:16.841 & -66:16:00.06 & $-45.45 \pm 0.10$ & $40.45 \pm 0.10$ & $12.28 \pm 0.06$ & 99.9\\
10164923-6603058\tablenotemark{d} & (M5) & 10:16:49.104 & -66:03:05.37 & $-47.00 \pm 0.12$ & $38.49 \pm 0.13$ & $11.84 \pm 0.07$ & 99.8\\
10222892-6115485 & (M5) & 10:22:28.822 & -61:15:48.06 & $-45.37 \pm 0.17$ & $30.11 \pm 0.16$ & $11.01 \pm 0.08$ & 95.5\\
\enddata
\tablenotetext{a}{Spectral types between parentheses are photometric spectral types were estimated from the absolute \emph{Gaia} $G$--band magnitude.\added{ Other spectral types reported in this table are from \cite{1975mcts.book.....H}.}}
\tablenotetext{b}{J2000 position at epoch 2015 from the \emph{Gaia}~DR2 catalog.}
\tablenotetext{c}{Bayesian probability for Volans-Carina membership obtained with BANYAN~$\Sigma$.}
\tablenotetext{d}{\emph{Gaia}~DR2 absolute $G$ magnitude versus $G - G_{\rm RP}$ color is consistent with an unresolved binary.}
\tablenotetext{e}{\emph{Gaia}~DR2 absolute $G$ magnitude versus $G - G_{\rm RP}$ color is consistent with an unresolved triple.}
\tablenotetext{f}{\emph{Gaia}~DR2 absolute $G$ magnitude versus $G - G_{\rm RP}$ color is consistent with an unresolved quadruple.}
\tablecomments{All objects with a parallax measurement have their proper motion and parallax measurements from \cite{Lindegren:2018gy}. All objects without parallax measurements have their proper motion measurements from a combination of 2MASS and AllWISE (e.g., see the method of \citealt{2015ApJ...798...73G}).}
\end{deluxetable*}

We used the empirical color correction displayed in Figure~\ref{fig:ple_corr} to correct all MIST isochrones and we built an age probability density function for both the Pleiades and Volans-Carina, using all members listed in Table~\ref{tab:phot} (but excluding 4 over-luminous members (HD~83359, J0917--6628, J0926--6236 and HD~309681) that are likely multiple systems, see Figure~\ref{fig:volans_ple_isoc}) with the method described by \cite{Gagne:2018un}. In summary, the minimum $N\sigma$ distance is calculated between a star and each corrected MIST model isochrone (excluding giant phases) in color-magnitude space, where $\sigma$ are the measurement errors for the star in color and absolute magnitude. The $N\sigma$ distance is translated to a probability density function by assuming that the measurement errors are Gaussian, and the probability density functions of all association members are multiplied to obtain a final probability density for the age of the association as a whole. In this analysis, we also included the error bars on the color correction in the $N\sigma$ calculations, which means that the uncertainty caused by the intrinsic spread of Pleiades members in this color-magnitude space is taken into account.

\begin{figure}
	\centering
	\includegraphics[width=0.465\textwidth]{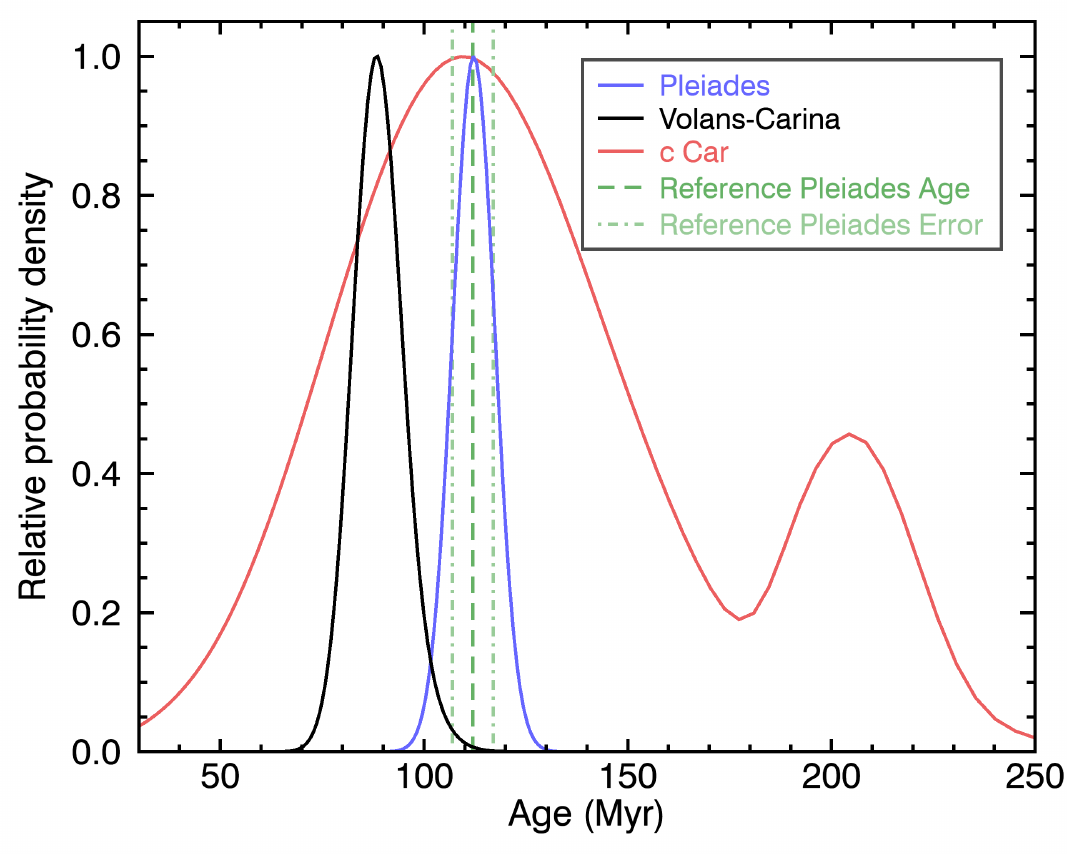}
	\caption{Age probability density function for Volans (black line) and the Pleiades (blue line) derived with our comparison to empirically corrected MIST isochrones. The age of the Pleiades ($112 \pm 5$\,Myr) that was assumed to correct the MIST isochrones is displayed with green dashed and dot-dashed lines. We find an age of \age\,Myr for Volans-Carina. See Section~\ref{sec:age} for more detail.}
	\label{fig:age_pdf}
\end{figure}

The resulting age probability density functions are displayed in Figure~\ref{fig:age_pdf} for the Pleiades and Volans-Carina. We find a Pleiades age of $105^{+13}_{-9}$\,Myr consistent with our initial assumption of $112 \pm 5$\,Myr, and we find an age of \age\,Myr for Volans-Carina. This age is consistent with the determination of \cite{2018arXiv180409058F} based on a \emph{Gaia}~DR1 absolute $G$ versus \emph{Gaia}--2MASS $G-J$ color-magnitude diagram and X-ray properties of its 8 members identified by \cite{2017AJ....153..257O}.\deleted{ The error bars are larger for the age of Volans-Carina because of its smaller number of members. We note that most of the age information relies on the most massive B8-type member c~Car, because it is located in a region of the color-magnitude diagram where isochrones vary strongly with age. Excluding c~Car from this analysis would result in a much wider 68\% confidence age range of 100--400\,Myr.} Future age determinations based on the lithium depletion boundary will be useful to confirm our age measurement. At the age of the Pleiades, the lithium depletion boundary is expected to happen at spectral types $\simeq$\,M6--M7 \citep{2015ApJ...813..108D}. There are 4 candidate members listed in Table~\ref{tab:cms} with estimated spectral types that fall in this range.

\begin{figure}
	\centering
	\includegraphics[width=0.465\textwidth]{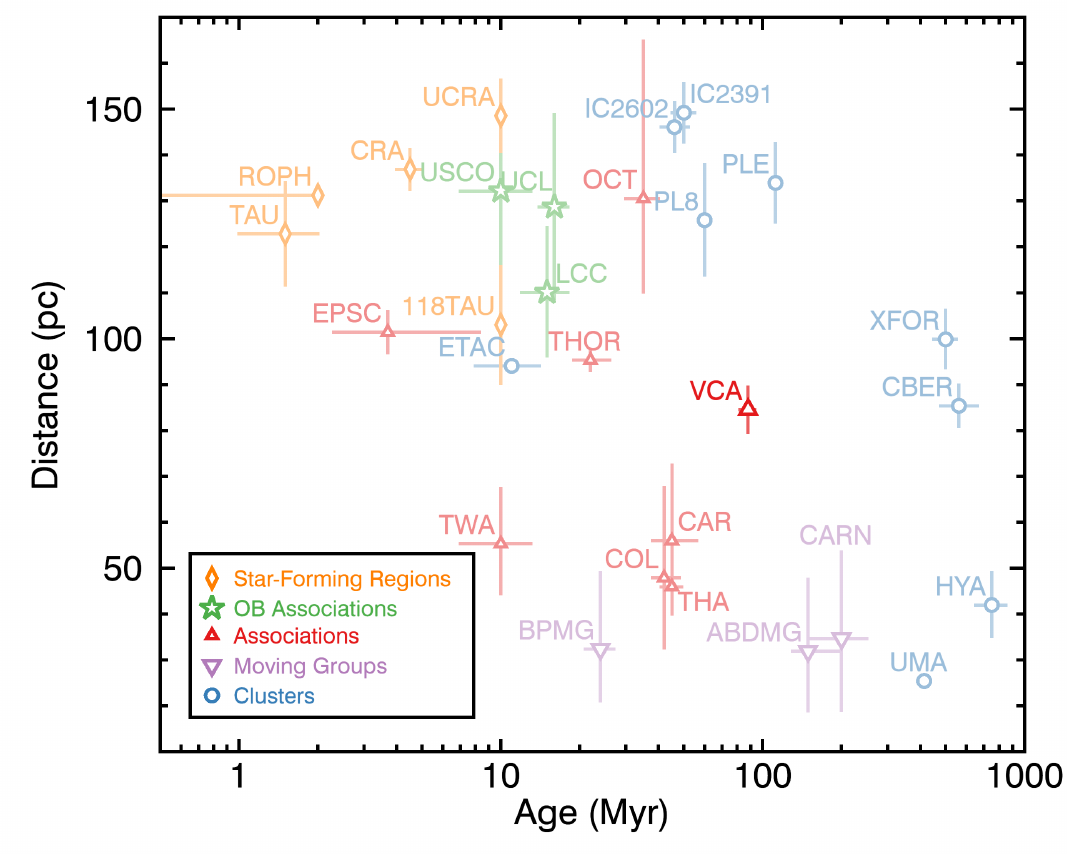}
	\caption{Distribution of ages and distances of young associations within 150\,pc of the Sun. The Volans-Carina Association (VCA) fills a region in age and distance unoccupied by any other association. The full names of young associations are: 118~Tau (118TAU), AB~Doradus (ABDMG), $\beta$~Pictoris ($\beta$PMG), Carina (CAR), Carina-Near (CARN), Coma Berenices (CBER), Columba (COL), Corona~Australis (CRA), $\epsilon$~Chamaeleontis (EPSC), $\eta$~Chamaeleontis (ETAC), the Hyades cluster (HYA), Lower Centaurus Crux (LCC), Octans (OCT), Platais~8 (PL8), the Pleiades cluster (PLE), $\rho$~Ophiuchi (ROPH), the Tucana-Horologium association (THA), 32~Orionis (THOR), TW~Hya (TWA), Upper Centaurus Lupus (UCL), Upper~CrA (UCRA), the core of the Ursa~Major cluster (UMA), Upper~Scorpius (USCO), Taurus (TAU), Volans-Carina (VCA) and $\chi^1$~For (XFOR). See Section~\ref{sec:age} for more detail.}
	\label{fig:agedist}
\end{figure}

In Figure~\ref{fig:agedist}, we show the global properties of Volans-Carina compared to other young associations within 150\,pc of the Sun. Volans-Carina fills a space un-occupied by any other association in the age--distance plane, which presents the opportunity to study a $\simeq$\,90\,Myr-old population brighter than the Pleiades, but not nearby enough that its members are spread on a large region of the sky.

\subsection{Chromospheric Activity}\label{sec:activ}

Young stars have stronger chromospheric activity compared to their field-aged counterparts, which translates into stronger and more variable H$\alpha$ emission (e.g., \citealt{2007AJ....133.2258S}), more frequent flares (e.g., \citealt{2010AJ....140.1402H}) and stronger emission at UV and X-ray wavelengths (e.g., \citealp{2003ApJ...585..878K,2011ApJ...727...62R,2013ApJ...774..101R,2014ApJ...788...81M}). This stronger level of chromospheric activity is correlated with faster rotation rates (e.g., see \citealt{2010AA...520A..15M}), as rotation drives stellar dynamos (e.g., \citealt{charbonneau2010,2012AJ....143...93R}).

\begin{figure}
	\centering
	\includegraphics[width=0.465\textwidth]{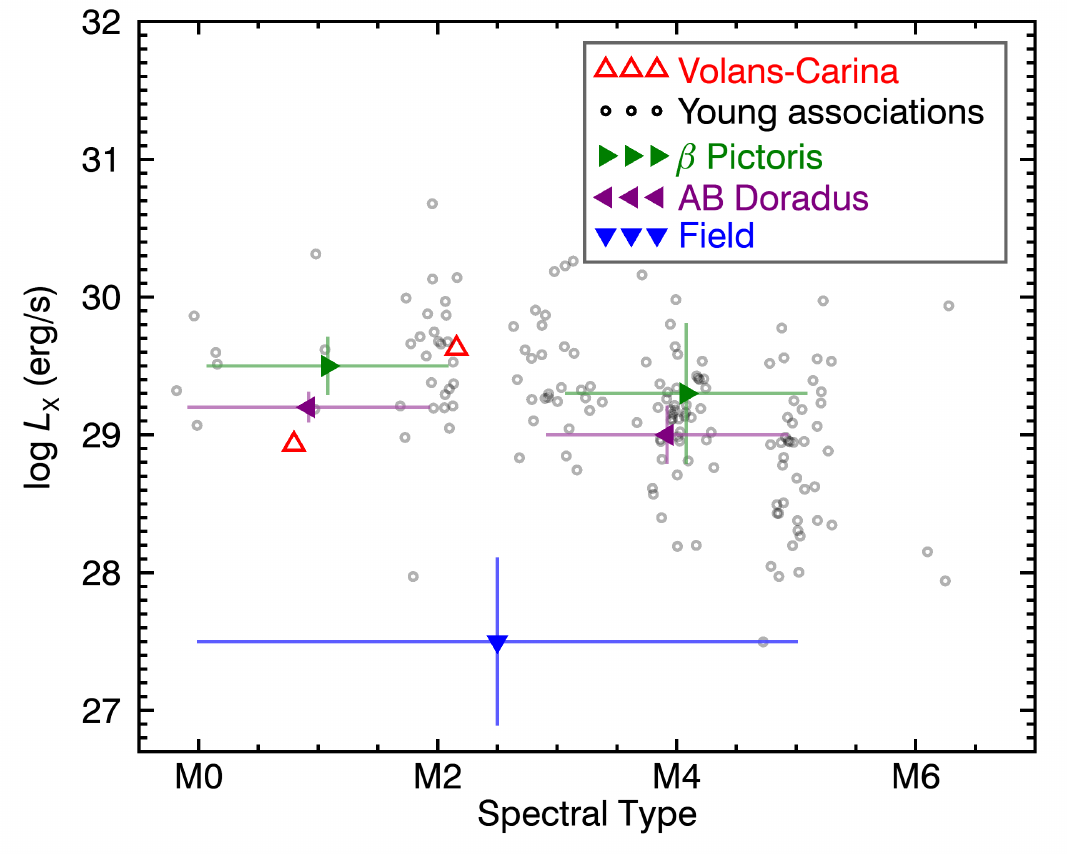}
	\caption{\emph{ROSAT} X-ray luminosity of Volans-Carina M-type candidate members compared to the \cite{2014ApJ...788...81M} compilations and the new candidate members of young associations identified by \cite{2018ApJ...862..138G}. The two Volans-Carina objects detected in \emph{ROSAT} have a level of X-ray emission consistent with the age derived here. See Section~\ref{sec:activ} for more detail.}
	\label{fig:xray}
\end{figure}

\begin{figure}
	\centering
	\includegraphics[width=0.465\textwidth]{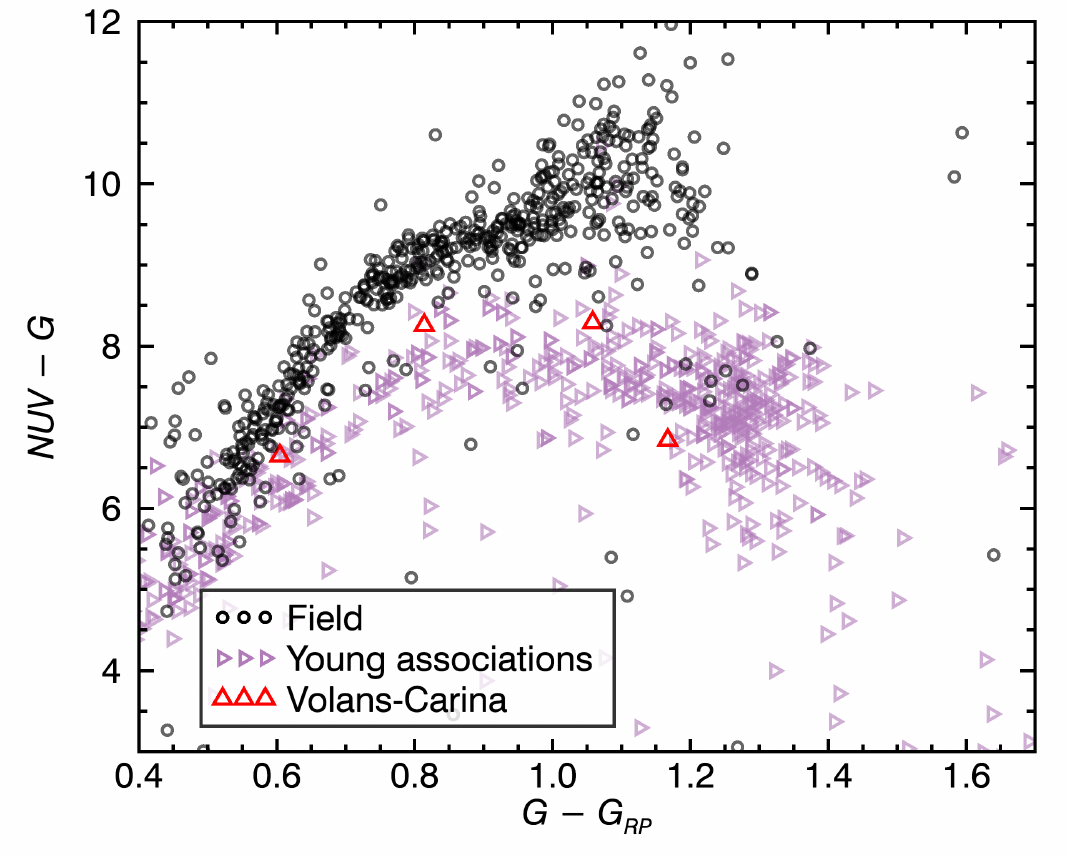}
	\caption{\emph{GALEX}--\emph{Gaia} $NUV - G$ color as a function of \emph{Gaia} $G - G_{\rm RP}$ colors of field stars (black circles), members of young associations (compiled by  \citealt{2018ApJ...856...23G}; purple leftward triangles) and candidate members of Volans-Carina detected in \emph{GALEX} (red upward triangles). None of the new candidate members of Volans-Carina listed in Table~\ref{tab:cms} are detected in \emph{GALEX}. The Volans-Carina candidate members have $NUV - G$ colors bluer than the field sequence, which is indicative of strong chromospheric activity and consistent with a young age. See Section~\ref{sec:activ} for more detail.}
	\label{fig:nuv}
\end{figure}

We cross-matched our sample of members and candidate members with the 
\emph{ROSAT} all-sky survey \citep{2016AA...588A.103B} and the \emph{GALEX} catalog \citep{2005ApJ...619L...1M} and found that 5 of them have entries in one or both of the catalogs. We compared the X-ray and UV emission properties of these stars with those of field stars and members of young associations compiled by \cite{2018ApJ...856...23G} in Figures~\ref{fig:xray} and \ref{fig:nuv}. All 5 stars have emission levels typical for ages younger than a few hundred Myr, consistent with our age determination in Section~\ref{sec:age}.


\subsection{Present-Day Mass Function}\label{sec:imf}

In this section we construct a preliminary present-day mass function for the Volans-Carina association, based only on the members with a parallax mesurement in \emph{Gaia}~DR2. The analysis presented here does not include errors on the model-derived masses or small number statistics Poisson error bars -- A more detailed measurement will be done after a spectroscopic confirmation of all Volans-Carina members presented here.

We calculated a preliminary present-day mass function of the Volans-Carina association by comparing the 58 candidate members in Table~\ref{tab:phot} with the empirically corrected MIST isochrone track discussed in Section~\ref{sec:age}. The mass of each star was assigned with the point on the model track at the closest $N\sigma$ separation in a $G$ versus $G-G_{\rm RP}$ color-magnitude diagram. This $N\sigma$ separation is calculated using the error of the $G - G_{\rm RP}$ color and the absolute $G$-band measurement, as well as the error on the empirical color correction discussed in Section~\ref{sec:age}. The eight candidates with colors redder than the full MIST sequence were ignored, and they likely correspond to stars with masses below 0.1\,\msol. Their spectral types are in the range M6--M9.\added{ We refrain from using less reliable substellar cooling tracks to estimate the masses of these redder objects until a deeper survey focused on the Volans-Carina brown dwarfs allows us to uncover additional such low-mass objects.}

\begin{figure}
	\centering
	\includegraphics[width=0.465\textwidth]{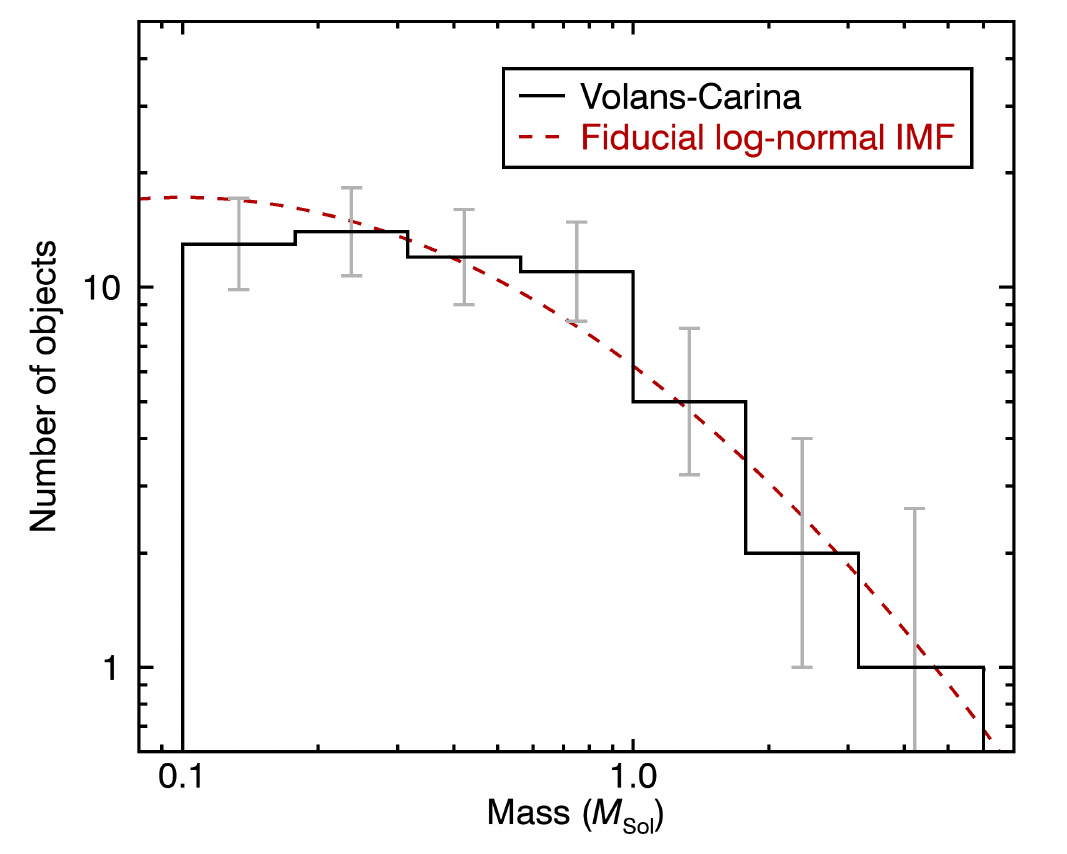}
	\caption{Preliminary present-day mass function for the Volans-Carina members with a \emph{Gaia}~DR2 parallax (black lines), compared with a fiducial log-normal initial mass function ($\sigma = 0.5$\,dex, $m_c = 0.25$\,\msol) anchored on the 1--3\,\msol\ members. The current census of Volans-Carina members seems complete down to $\simeq$\,0.2\,\msol. See Section~\ref{sec:imf} for more detail.}
	\label{fig:imf}
\end{figure}

The resulting present-day mass function is displayed in Figure~\ref{fig:imf}, with a fiducial log-normal initial mass function ($\sigma = 0.5$\,dex, $m_c = 0.25$\,\msol; \citealt{2010AJ....139.2679B}) anchored on the members with masses 1--3\,\msol. Our preliminary present-day mass function is well represented by a log-normal distribution in the range $\geq$\,0.2\,\msol, indicating that our census of Volans-Carina members is likely near completion in this range of masses.

If we assume that the fiducial log-normal present-day mass function remains valid down to the lowest-mass brown dwarfs, we estimate a total stellar and substellar population of $\simeq$\,120 members in Volans-Carina. Counting all candidate members not in \emph{Gaia}~DR2 or otherwise excluded from our determination of the present-day mass function, around $\simeq$\,40 members would remain to be discovered, $\simeq$\,75\% of which would be in the substellar regime. These missing objects are likely too faint to be in \emph{Gaia}~DR2.

\subsection{A White Dwarf Interloper}\label{sec:wd}

An inspection of the \emph{Gaia}~DR2 entries in the vicinity of Volans-Carina that share similar proper motions revealed one potential white dwarf member (Gaia~DR2~5298305642425782528, J0852--6143 hereafter), located at J2000 coordinates 8$^{\rm h}$52$^{\rm m}$40$^{\rm s}$.39, --61\textdegree43$'$46\farcs4 at epoch 2015. However, this object only obtains a 19.7\% Volans-Carina membership probability based on BANYAN~$\Sigma$.

\begin{figure}
	\centering
	\includegraphics[width=0.465\textwidth]{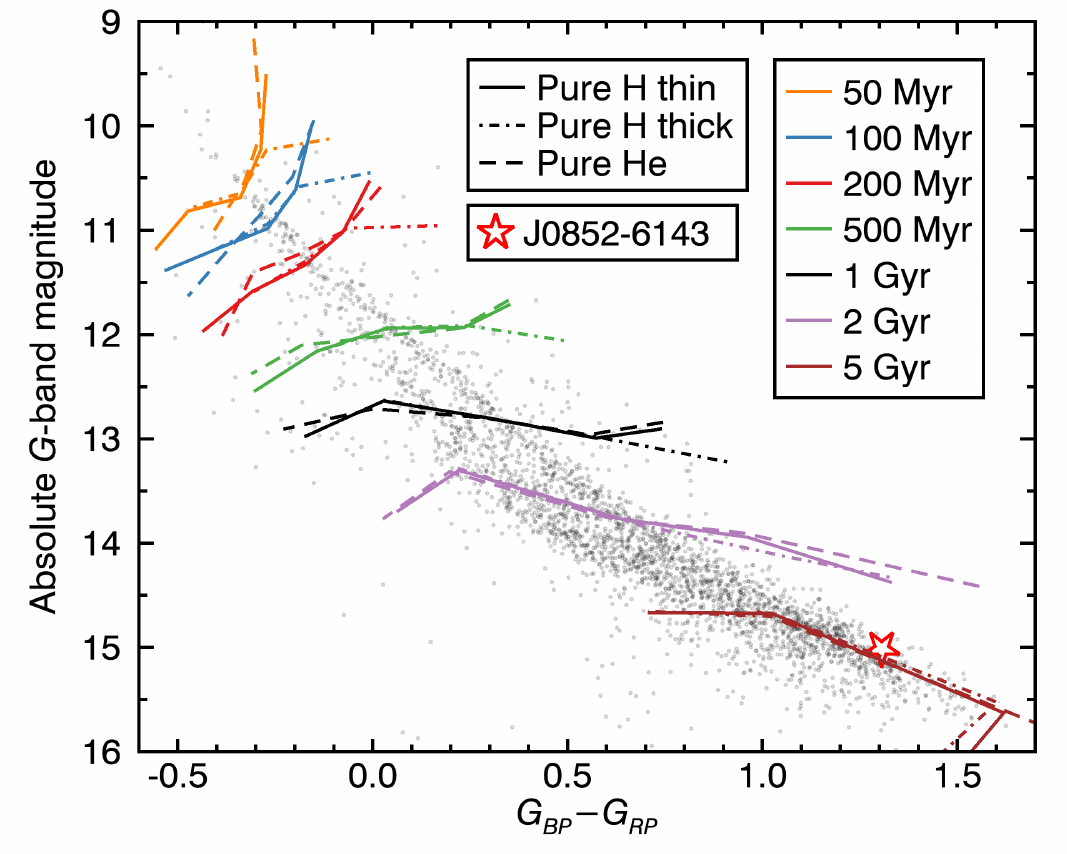}
	\caption{Color-magnitude position of the white dwarf J0852--6143 (red star) compared to the 100\,pc sample of \emph{Gaia}~DR2 white dwarfs (black circles) and isochrones calculated from the cooling tracks of \cite{2001PASP..113..409F}. J0852--6143 is older than 3\,Gyr, regardless of its atmospheric composition. See Section~\ref{sec:wd} for more detail.}
	\label{fig:volans_wd}
\end{figure}

We compared J0852--6143 with the \cite{2001PASP..113..409F} white dwarf cooling tracks\footnote{Available at \url{http://www.astro.umontreal.ca/~bergeron/CoolingModels/}} (see also \citealp{2006AJ....132.1221H,2006ApJ...651L.137K,2011ApJ...730..128T,2011ApJ...737...28B}) in a \emph{Gaia}~DR2 absolute $G$ magnitude versus $G_{\rm BP} - G_{\rm RP}$ color-magnitude diagram (see Figure~\ref{fig:volans_wd}) to construct a probability density function of its mass, age and temperature. We used unit priors on all parameters, and compared J0852--6143 to the models with pure hydrogen (thin and thick) and pure helium atmospheres.

We calculated its most probable age, \teff\ and mass by marginalizing over all other dimensions (including the atmospheric composition), and found an age of 3--8\,Gyr, a mass of $\simeq$\,0.15\,\msol\ and \teff\,$=$\,$3000_{-800}^{+500}$\,K. The extremely low mass of this white dwarf is also indicative that it is likely a more massive, older and slightly cooler double-degenerate binary. Given that the age of J0852--6143 is safely much older than that of Volans-Carina, we reject it as a candidate member. \added{Based on our selection criteria and the Monte Carlo analysis of \citeauthor{2018ApJ...856...23G} (\citeyear{2018ApJ...856...23G}; Section~8), we expect less than one additional such interloper in our list of members and candidate members presented in Table~\ref{tab:cms}.}

\subsection{Is Volans-Carina Related to Other Known Associations?}\label{sec:relat}

Despite its similar age to the Pleiades association and the AB~Doradus moving group, the average $UVW$ space velocities of the Volans-Carina members with full kinematics ($U = -16.1 \pm 0.9$\,\kms, $V = -28.4 \pm 1.1$\,\kms, $W = -1.1 \pm 1.2$\,\kms) are different from those of the Pleiades ($U = -6.4 \pm 1.9$\,\kms, $V = -28.7 \pm 0.7$\,\kms, $W = -14.1 \pm 1.1$\,\kms) and the AB~Doradus moving group ($U = -7.7 \pm 0.8$\,\kms, $V = -27.8 \pm 0.7$\,\kms, $W = -15.3 \pm 1.8$\,\kms) in the $U$ and $W$ directions. \added{Both the spatial and kinematic sizes of Volans-Carina (reported in Table~\ref{tab:vca}) are typical of other nearby associations such as the Pleiades and Hyades (see panels c and d of Figure 3 in \citealt{2018ApJ...856...23G}). Its spatial size is slightly smaller than the typical sizes of the looser nearby young moving groups such as Tucana-Horologium, $\beta$~Pictoris and AB~Doradus ($\sim$\,10--20\,pc) although its kinematic size is comparable to them.}

Comparing the center of the BANYAN~$\Sigma$ Volans-Carina model in $UVW$ space to those of the other 27 associations included in the algorithm (see \citealt{2018ApJ...856...23G}), we find that the closest known association in $UVW$ space is the slightly younger ($\simeq$\,60\,Myr; \citealt{1998AJ....116.2423P}) Platais~8\footnote{The members of Platais~8 have average $UVW$ space velocities of $U = -11.5 \pm 1.0$\,\kms, $V = -23.2 \pm 0.8$\,\kms, $W = -4.1 \pm 0.5$\,\kms.}, located at 8\,\kms\ from Volans-Carina in $UVW$ space, and $\simeq$\,46\,pc in $XYZ$ space. Both associations are spatially much smaller than the physical separation between the two groups (5.0\,pc and \spatsize\,pc respectively for Platais~8 and Volans-Carina), but their relative proximity and similar ages and kinematics indicates that the two groups may correspond to two star-formation events that originally took place in the same complex of molecular clouds.

\section{CONCLUSIONS}\label{sec:conclusion}

We present a characterization of the young Volans-Carina association, identified as `Group~30', a group of 8 co-moving stars by \cite{2017AJ....153..257O} and \cite{2018arXiv180409058F}. We find 11 additional high-likelihood members with \emph{Gaia}~DR2, and we use the MIST isochrones anchored on the Pleiades to determine an age of \age\,Myr for Volans-Carina. We build a kinematic model of its members in $XYZUVW$ space and include it in the BANYAN~$\Sigma$ Bayesian classification tool to identify \gaiacandidatemembers\ additional candidate members with estimated spectral types in the range A0--M9 in \emph{Gaia}~DR2, and \tmwisecandidatemembers\ with estimated spectral types in the range M6--M9 in 2MASS and AllWISE.

We find that the present-day mass function of Volans-Carina follows a fiducial log-normal distribution and seems complete for masses above 0.2\,\msol, and we estimate that $\simeq$\,40 additional members, mostly brown dwarfs, remain to be discovered in the association. Brown dwarfs in Volans-Carina with spectral types $\gtrsim$\,L1 are too faint to be detected in 2MASS, but the members with spectral types up to $\gtrsim$\,L6 should be detected in the WISE $W2$ band. Proper motion surveys based on only WISE such as Backyard Worlds: Planet~9 \citep{2017ApJ...841L..19K} will be needed to identify them.

\acknowledgments

We thank the anonymous referee for valuable suggestions. We thank Ricky L. Smart, Pierre Bergeron and Am\'elie Simon for useful comments, and Pierre Bergeron for sending white dwarf model tracks in the \emph{Gaia}~DR2 passbands. This research made use of: the SIMBAD database and VizieR catalog access tool, operated at the Centre de Donn\'ees astronomiques de Strasbourg, France \citep{2000AAS..143...23O}; data products from the Two Micron All Sky Survey (\emph{2MASS}; \citealp{2006AJ....131.1163S}), which is a joint project of the University of Massachusetts and the Infrared Processing and Analysis Center (IPAC)/California Institute of Technology (Caltech), funded by the National Aeronautics and Space Administration (NASA) and the National Science Foundation \citep{2006AJ....131.1163S}; data products from the \emph{Wide-field Infrared Survey Explorer} (\emph{WISE}; and \citealp{2010AJ....140.1868W}), which is a joint project of the University of California, Los Angeles, and the Jet Propulsion Laboratory (JPL)/Caltech, funded by NASA. The Digitized Sky Surveys were produced at the Space Telescope Science Institute under U.S. Government grant NAG W-2166. The images of these surveys are based on photographic data obtained using the Oschin Schmidt Telescope on Palomar Mountain and the UK Schmidt Telescope. The plates were processed into the present compressed digital form with the permission of these institutions. The Second Palomar Observatory Sky Survey (POSS-II) was made by the California Institute of Technology with funds from the National Science Foundation, the National Geographic Society, the Sloan Foundation, the Samuel Oschin Foundation, and the Eastman Kodak Corporation. The Oschin Schmidt Telescope is operated by the California Institute of Technology and Palomar Observatory. This work presents results from the European Space Agency (ESA) space mission Gaia. Gaia data are being processed by the Gaia Data Processing and Analysis Consortium (DPAC). Funding for the DPAC is provided by national institutions, in particular the institutions participating in the Gaia MultiLateral Agreement (MLA). The Gaia mission website is https://www.cosmos.esa.int/gaia. The Gaia archive website is https://archives.esac.esa.int/gaia. This research was started at the NYC Gaia DR2 Workshop at the Center for Computational Astrophysics of the Flatiron Institute in 2018 April. Part of this research was carried out at the Jet Propulsion Laboratory, California Institute of Technology, under a contract with the National
Aeronautics and Space Administration.  EEM acknowledges support from the NASA NExSS program.

\startlongtable
 \begin{deluxetable*}{llcccccc}
\tablecolumns{8}
\tablecaption{Members and Candidates of Volans-Carina used for Isochrone Age Determination.\label{tab:phot}}
 \tablehead{\colhead{Name} & \colhead{Spectral} & \colhead{R.A.\tablenotemark{b}} & \colhead{Decl.\tablenotemark{b}} & \colhead{Trig. Dist.} & \colhead{$G$} & \colhead{$G_{\rm RP}$} & \colhead{$G_{\rm BP}$}
 \\
 \colhead{} & \colhead{Type\tablenotemark{a}} & \colhead{(hh:mm:ss.sss)} & \colhead{(dd:mm:ss.ss)} & \colhead{(pc)} & \colhead{(mag)} & \colhead{(mag)} & \colhead{(mag)}
}
\startdata
TYC 8933-327-1 & (K4) & 08:26:49.603 & -63:46:36.08 & $76.5 \pm 0.6$ & $10.5379 \pm 0.0005$ & $9.933 \pm 0.001$ & $11.008 \pm 0.001$\\
J0838-6716 & (M5) & 08:38:33.458 & -67:16:35.95 & $85.7 \pm 0.6$ & $17.050 \pm 0.001$ & $15.703 \pm 0.003$ & $19.07 \pm 0.03$\\
J0848-6113 & (M5) & 08:48:55.579 & -61:13:25.38 & $87.5 \pm 0.6$ & $16.795 \pm 0.001$ & $15.482 \pm 0.004$ & $18.68 \pm 0.02$\\
J0851-7140 & (K8) & 08:51:15.710 & -71:40:14.37 & $79.3 \pm 0.1$ & $11.5974 \pm 0.0007$ & $10.818 \pm 0.002$ & $12.289 \pm 0.003$\\
J0852-6552 & (M4) & 08:52:40.594 & -65:52:27.06 & $89.0 \pm 0.5$ & $15.132 \pm 0.002$ & $13.869 \pm 0.004$ & $16.821 \pm 0.009$\\
J0852-6401 & (M6) & 08:52:41.501 & -64:01:22.14 & $83.1 \pm 0.6$ & $17.227 \pm 0.001$ & $15.849 \pm 0.003$ & $19.37 \pm 0.03$\\
J0854-7055 & (M5) & 08:54:45.629 & -70:55:07.08 & $81.2 \pm 0.4$ & $15.188 \pm 0.001$ & $13.948 \pm 0.002$ & $16.80 \pm 0.01$\\
c Car & B8 II & 08:55:02.791 & -60:38:39.00 & $94 \pm 3$ & $3.728 \pm 0.005$ & $3.889 \pm 0.003$ & $3.776 \pm 0.005$\\
J0855-6146 & (M5) & 08:55:56.496 & -61:46:05.11 & $93.0 \pm 0.6$ & $15.936 \pm 0.001$ & $14.679 \pm 0.002$ & $17.660 \pm 0.009$\\
J0859-6919 & (M2) & 08:59:13.574 & -69:19:43.48 & $87.1 \pm 0.6$ & $12.7541 \pm 0.0004$ & $11.792 \pm 0.001$ & $13.693 \pm 0.002$\\
J0859-6918 & (M2) & 08:59:15.199 & -69:18:41.80 & $87.3 \pm 0.2$ & $12.812 \pm 0.001$ & $11.872 \pm 0.003$ & $13.741 \pm 0.004$\\
J0904-6325 & (M5) & 09:04:06.799 & -63:25:32.92 & $78.1 \pm 0.3$ & $15.329 \pm 0.001$ & $14.092 \pm 0.002$ & $16.927 \pm 0.006$\\
J0909-6826 & (M5) & 09:09:00.787 & -68:26:01.52 & $75 \pm 4$ & $15.716 \pm 0.002$ & $14.391 \pm 0.002$ & $17.469 \pm 0.008$\\
J0911-6631 & (M4) & 09:11:14.359 & -66:31:38.18 & $86.3 \pm 0.3$ & $14.6815 \pm 0.0006$ & $13.499 \pm 0.002$ & $16.141 \pm 0.005$\\
J0913-6522 & (M5) & 09:13:36.286 & -65:22:10.77 & $88.0 \pm 0.4$ & $15.7496 \pm 0.0008$ & $14.487 \pm 0.002$ & $17.43 \pm 0.01$\\
HD 80563 & F3V & 09:17:34.075 & -63:23:13.88 & $91.4 \pm 0.2$ & $8.2152 \pm 0.0004$ & $7.880 \pm 0.002$ & $8.451 \pm 0.001$\\
J0917-6628 & (M2) & 09:17:49.649 & -66:28:13.45 & $94.1 \pm 0.3$ & $13.076 \pm 0.001$ & $12.018 \pm 0.002$ & $14.215 \pm 0.005$\\
J0918-6310 & (M5) & 09:18:15.672 & -63:10:53.32 & $86.2 \pm 0.4$ & $15.933 \pm 0.001$ & $14.664 \pm 0.002$ & $17.67 \pm 0.01$\\
J0919-6640 & (M6) & 09:19:11.796 & -66:40:11.47 & $74.8 \pm 0.7$ & $17.443 \pm 0.001$ & $15.950 \pm 0.005$ & $20.18 \pm 0.05$\\
J0920-6607 & (K8) & 09:20:19.783 & -66:07:41.06 & $87.4 \pm 0.2$ & $12.025 \pm 0.002$ & $11.212 \pm 0.004$ & $12.762 \pm 0.006$\\
J0920-6329 & (M7) & 09:20:26.126 & -63:29:54.15 & $91 \pm 2$ & $18.524 \pm 0.002$ & $17.002 \pm 0.007$ & $20.7 \pm 0.1$\\
J0920-6309 & (M2) & 09:20:36.427 & -63:09:59.14 & $91.6 \pm 0.2$ & $13.111 \pm 0.002$ & $12.136 \pm 0.004$ & $14.099 \pm 0.007$\\
J0922-6756 & (M4) & 09:22:19.841 & -67:56:51.15 & $85.5 \pm 0.3$ & $14.518 \pm 0.002$ & $13.351 \pm 0.007$ & $15.916 \pm 0.008$\\
J0922-6206 & (M5) & 09:22:31.037 & -62:06:06.57 & $89.6 \pm 0.7$ & $17.067 \pm 0.001$ & $15.653 \pm 0.003$ & $19.35 \pm 0.03$\\
J0923-6556 & (M4) & 09:23:41.916 & -65:56:50.53 & $85.8 \pm 0.2$ & $14.6577 \pm 0.0005$ & $13.482 \pm 0.002$ & $16.092 \pm 0.004$\\
J0924-6254 & (M5) & 09:24:42.943 & -62:54:46.14 & $90.3 \pm 0.6$ & $16.650 \pm 0.001$ & $15.352 \pm 0.002$ & $18.50 \pm 0.01$\\
J0924-6856 & (M5) & 09:24:43.270 & -68:56:21.48 & $86.7 \pm 0.5$ & $15.9008 \pm 0.0009$ & $14.645 \pm 0.002$ & $17.568 \pm 0.006$\\
J0924-6216 & (M7) & 09:24:49.519 & -62:16:31.28 & $88 \pm 1$ & $18.465 \pm 0.002$ & $16.973 \pm 0.005$ & $20.9 \pm 0.1$\\
HD 309681 & G0 & 09:25:01.560 & -64:37:30.49 & $86.7 \pm 0.9$ & $9.026 \pm 0.001$ & $8.554 \pm 0.003$ & $9.375 \pm 0.003$\\
J0926-6236 & (M1) & 09:26:31.370 & -62:36:22.98 & $90.1 \pm 0.9$ & $12.6322 \pm 0.0009$ & $11.633 \pm 0.002$ & $13.645 \pm 0.003$\\
J0928-6553 & (M5) & 09:28:08.165 & -65:53:58.23 & $94.8 \pm 0.8$ & $16.801 \pm 0.001$ & $15.494 \pm 0.003$ & $18.67 \pm 0.02$\\
HD 82406 & A0V & 09:28:30.456 & -66:42:05.99 & $84 \pm 1$ & $5.8724 \pm 0.0007$ & $5.875 \pm 0.004$ & $5.905 \pm 0.003$\\
J0929-6345 & (M4) & 09:29:31.147 & -63:45:39.03 & $88.3 \pm 0.3$ & $14.7788 \pm 0.0007$ & $13.50 \pm 0.02$ & $16.202 \pm 0.008$\\
J0931-6414 & (M2) & 09:31:16.178 & -64:14:26.44 & $88.1 \pm 0.3$ & $12.859 \pm 0.001$ & $11.899 \pm 0.002$ & $13.819 \pm 0.004$\\
J0931-6419 & (M5) & 09:31:21.845 & -64:19:23.40 & $96.7 \pm 0.7$ & $16.502 \pm 0.001$ & $15.186 \pm 0.003$ & $18.33 \pm 0.02$\\
J0931-6600 & (M5) & 09:31:43.946 & -66:00:53.09 & $73.9 \pm 0.3$ & $20.61 \pm 0.03$ & $18.64 \pm 0.04$ & $21.0 \pm 0.6$\\
J0932-6908 & (M5) & 09:32:33.149 & -69:08:43.18 & $86.7 \pm 0.6$ & $15.599 \pm 0.001$ & $14.273 \pm 0.007$ & $16.94 \pm 0.02$\\
J0933-6251 & (M1) & 09:33:00.082 & -62:51:20.17 & $88.6 \pm 0.2$ & $12.5210 \pm 0.0009$ & $11.627 \pm 0.003$ & $13.345 \pm 0.004$\\
J0933-6330 & (M4) & 09:33:00.187 & -63:30:37.54 & $88.6 \pm 0.3$ & $14.1925 \pm 0.0006$ & $13.109 \pm 0.001$ & $15.399 \pm 0.003$\\
J0934-6500 & K7 & 09:34:55.690 & -65:00:07.10 & $86.8 \pm 0.2$ & $12.2903 \pm 0.0008$ & $11.431 \pm 0.004$ & $13.089 \pm 0.005$\\
HD 83359 & F5V & 09:34:56.376 & -64:59:57.38 & $87.4 \pm 0.2$ & $7.9007 \pm 0.0004$ & $7.537 \pm 0.002$ & $8.162 \pm 0.002$\\
HD 83523 & A2V & 09:36:05.098 & -64:57:00.25 & $87.8 \pm 0.3$ & $6.5449 \pm 0.0004$ & $6.484 \pm 0.004$ & $6.595 \pm 0.003$\\
HD 83948 & A9IV/V & 09:38:45.130 & -66:51:31.99 & $85.5 \pm 0.2$ & $7.3311 \pm 0.0003$ & $7.122 \pm 0.002$ & $7.487 \pm 0.002$\\
HD 83946 & F5V & 09:38:54.019 & -64:59:26.10 & $90.5 \pm 0.2$ & $8.7190 \pm 0.0005$ & $8.329 \pm 0.001$ & $8.994 \pm 0.001$\\
TYC 8953-1289-1 & (K4) & 09:39:10.366 & -66:46:14.85 & $84.3 \pm 0.2$ & $10.7231 \pm 0.0005$ & $10.094 \pm 0.001$ & $11.225 \pm 0.002$\\
J0941-6658 & (M5) & 09:41:06.749 & -66:58:57.79 & $95.8 \pm 0.7$ & $15.7639 \pm 0.0009$ & $14.476 \pm 0.002$ & $17.533 \pm 0.007$\\
J0943-6422 & (M5) & 09:43:00.014 & -64:22:28.53 & $95.7 \pm 0.6$ & $16.033 \pm 0.001$ & $14.718 \pm 0.004$ & $17.63 \pm 0.03$\\
J0945-6908 & (M3) & 09:45:15.134 & -69:08:22.66 & $83.2 \pm 0.1$ & $13.342 \pm 0.001$ & $12.309 \pm 0.003$ & $14.431 \pm 0.005$\\
TYC 8950-1447-1 & (K6) & 09:48:19.150 & -64:03:21.12 & $77.2 \pm 0.1$ & $11.034 \pm 0.001$ & $10.335 \pm 0.002$ & $11.627 \pm 0.003$\\
J0952-6731 & (M5) & 09:52:24.343 & -67:31:12.96 & $87.1 \pm 0.6$ & $15.6347 \pm 0.0007$ & $14.344 \pm 0.002$ & $17.425 \pm 0.007$\\
J1001-6508 & (M4) & 10:01:48.014 & -65:08:42.82 & $83.6 \pm 0.3$ & $15.019 \pm 0.001$ & $13.799 \pm 0.003$ & $16.587 \pm 0.007$\\
J1002-6845 & (M4) & 10:02:41.256 & -68:45:17.33 & $93.7 \pm 0.3$ & $14.427 \pm 0.001$ & $13.274 \pm 0.004$ & $15.783 \pm 0.007$\\
TYC 9210-1818-1 & (K6) & 10:08:21.890 & -67:56:39.63 & $86.8 \pm 0.2$ & $11.2997 \pm 0.0007$ & $10.590 \pm 0.001$ & $11.905 \pm 0.002$\\
J1008-6757 & (M5) & 10:08:21.938 & -67:57:03.45 & $87.3 \pm 0.5$ & $16.160 \pm 0.002$ & $14.863 \pm 0.005$ & $17.93 \pm 0.02$\\
J1010-6616 & (M5) & 10:10:16.841 & -66:16:00.06 & $81.5 \pm 0.4$ & $15.2555 \pm 0.0005$ & $13.962 \pm 0.002$ & $17.070 \pm 0.004$\\
CD-67 852 & (K0) & 10:12:29.772 & -67:52:32.79 & $81.9 \pm 0.2$ & $9.7674 \pm 0.0007$ & $9.247 \pm 0.003$ & $10.157 \pm 0.003$\\
J1016-6603 & (M5) & 10:16:49.104 & -66:03:05.37 & $84.4 \pm 0.5$ & $15.2210 \pm 0.0006$ & $13.950 \pm 0.002$ & $16.935 \pm 0.005$\\
J1022-6115 & (M5) & 10:22:28.822 & -61:15:48.06 & $90.9 \pm 0.7$ & $16.718 \pm 0.001$ & $15.396 \pm 0.002$ & $18.55 \pm 0.04$\\
J1022-5847 & (M2) & 10:22:38.585 & -58:47:59.67 & $82.8 \pm 0.2$ & $12.8032 \pm 0.0005$ & $11.846 \pm 0.002$ & $13.767 \pm 0.003$\\
\enddata
\tablenotetext{a}{Spectral types between parentheses were estimated from the absolute \emph{Gaia} $G$--band magnitude.}
\tablenotetext{b}{J2000 position at epoch 2015 from the \emph{Gaia}~DR2 catalog.}
\tablecomments{All measurements in this table are from \cite{Lindegren:2018gy}, except for spectral types for which references are listed in Table~\ref{tab:orig}.}
\end{deluxetable*}

\emph{JG} wrote the codes, manuscript, generated figures and led all analysis; \emph{JKF} provided data on the \cite{2017AJ....153..257O} members and general comments; \emph{EEM} provided general comments, helped parse the literature and provide additional information on the members of Volans-Carina, and provided help with the determination of the association name and reddening.

\software{BANYAN~$\Sigma$ \citep{2018ApJ...856...23G}.}

\bibliographystyle{apj}

\begin{thebibliography}{}
\expandafter\ifx\csname natexlab\endcsname\relax\def\natexlab#1{#1}\fi

\bibitem[{Ammons {et~al.}(2006)Ammons, Robinson, Strader, Laughlin, Fischer, \&
  Wolf}]{2006ApJ...638.1004A}
Ammons, S.~M., Robinson, S.~E., Strader, J., {et~al.} 2006, The Astrophysical
  Journal, 638, 1004

\bibitem[{Andrews {et~al.}(2017)Andrews, Chanam{\'e}, \&
  Ag{\"u}eros}]{2017MNRAS.472..675A}
Andrews, J.~J., Chanam{\'e}, J., \& Ag{\"u}eros, M.~A. 2017, Monthly Notices of
  the Royal Astronomical Society, 472, 675

\bibitem[{Babusiaux {et~al.}(2018)Babusiaux, van Leeuwen, Barstow, Jordi,
  Vallenari, Bossini, Bressan, {Gaia Collaboration}, \& {DPAC
  co-authors}}]{Babusiaux:2018di}
Babusiaux, C., van Leeuwen, F., Barstow, M., {et~al.} 2018, Astronomy {\&}
  Astrophysics, doi:10.1051/0004-6361/201832843

\bibitem[{Bell {et~al.}(2015)Bell, Mamajek, \& Naylor}]{2015MNRAS.454..593B}
Bell, C. P.~M., Mamajek, E.~E., \& Naylor, T. 2015, Monthly Notices of the
  Royal Astronomical Society, 454, 593

\bibitem[{Bergeron {et~al.}(2011)Bergeron, Wesemael, Dufour, Beauchamp, Hunter,
  Saffer, Gianninas, Ruiz, Limoges, Dufour, Fontaine, \&
  Liebert}]{2011ApJ...737...28B}
Bergeron, P., Wesemael, F., Dufour, P., {et~al.} 2011, The Astrophysical
  Journal, 737, 28

\bibitem[{Bochanski {et~al.}(2010)Bochanski, Hawley, Covey, West, Reid,
  Golimowski, \& Ivezi{\'c}}]{2010AJ....139.2679B}
Bochanski, J.~J., Hawley, S.~L., Covey, K.~R., {et~al.} 2010, The Astronomical
  Journal, 139, 2679

\bibitem[{Boller {et~al.}(2016)Boller, Freyberg, Tr{\"u}mper, Haberl, Voges, \&
  Nandra}]{2016AA...588A.103B}
Boller, T., Freyberg, M.~J., Tr{\"u}mper, J., {et~al.} 2016, Astronomy {\&}
  Astrophysics, 588, A103

\bibitem[{Capitanio {et~al.}(2017)Capitanio, Lallement, Vergely, Elyajouri, \&
  Monreal-Ibero}]{2017AA...606A..65C}
Capitanio, L., Lallement, R., Vergely, J.~L., Elyajouri, M., \& Monreal-Ibero,
  A. 2017, Astronomy and Astrophysics, 606, A65

\bibitem[{Charbonneau(2010)}]{charbonneau2010}
Charbonneau, P. 2010, Living Reviews in Solar Physics, 7, 3

\bibitem[{Choi {et~al.}(2016)Choi, Dotter, Conroy, Cantiello, Paxton, \&
  Johnson}]{2016ApJ...823..102C}
Choi, J., Dotter, A., Conroy, C., {et~al.} 2016, The Astrophysical Journal,
  823, 102

\bibitem[{Cropper {et~al.}(2018)Cropper, Katz, Sartoretti, \& {et
  al.}}]{Cropper:2018jx}
Cropper, M., Katz, D., Sartoretti, P., \& {et al.} 2018, Astronomy {\&}
  Astrophysics, doi:10.1051/0004-6361/201832763

\bibitem[{Cucchiaro {et~al.}(1977)Cucchiaro, Macau-Hercot, Jaschek, \&
  Jaschek}]{1977AAS...30...71C}
Cucchiaro, A., Macau-Hercot, D., Jaschek, M., \& Jaschek, C. 1977, Astronomy
  {\&} Astrophysics Supplement Series, 30, 71

\bibitem[{Dahm(2015)}]{2015ApJ...813..108D}
Dahm, S.~E. 2015, The Astrophysical Journal, 813, 108

\bibitem[{de~Vaucouleurs(1957)}]{1957MNRAS.117..449D}
de~Vaucouleurs, A. 1957, Monthly Notices of the Royal Astronomical Society,
  117, 449

\bibitem[{Dupuy \& Liu(2012)}]{2012ApJS..201...19D}
Dupuy, T.~J., \& Liu, M.~C. 2012, The Astrophysical Journal Supplement, 201, 19

\bibitem[{Evans {et~al.}(2018{\natexlab{a}})Evans, Riello, De~Angeli, Carrasco,
  Montegriffo, Fabricius, Jordi, Palaversa, Diener, Busso, Cacciari, \& van
  Leeuwen}]{2018arXiv180409368E}
Evans, D.~W., Riello, M., De~Angeli, F., {et~al.} 2018{\natexlab{a}},
  arXiv.org, arXiv:1804.09368

\bibitem[{Evans {et~al.}(2018{\natexlab{b}})Evans, Riello, De~Angeli, Carrasco,
  Montegriffo, Fabricius, Jordi, Palaversa, Diener, Busso, Weiler, Cacciari, \&
  van Leeuwen}]{Evans:2018cj}
---. 2018{\natexlab{b}}, Astronomy {\&} Astrophysics,
  doi:10.1051/0004-6361/201832756

\bibitem[{Faherty {et~al.}(2018)Faherty, Bochanski, Gagn{\'e}, Nelson, Coker,
  Smithka, Desir, \& Vasquez}]{2018arXiv180409058F}
Faherty, J.~K., Bochanski, J.~J., Gagn{\'e}, J., {et~al.} 2018, arXiv.org,
  arXiv:1804.09058

\bibitem[{Faherty {et~al.}(2016)Faherty, Riedel, Cruz, Gagn{\'e}, Filippazzo,
  Lambrides, Fica, Weinberger, Thorstensen, Tinney, Baldassare, Lemonier, \&
  Rice}]{2016ApJS..225...10F}
Faherty, J.~K., Riedel, A.~R., Cruz, K.~K., {et~al.} 2016, The Astrophysical
  Journal Supplement Series, 225, 10

\bibitem[{Feiden(2016)}]{2016AA...593A..99F}
Feiden, G.~A. 2016, Astronomy and Astrophysics, 593, A99

\bibitem[{Fontaine {et~al.}(2001)Fontaine, Brassard, \&
  Bergeron}]{2001PASP..113..409F}
Fontaine, G., Brassard, P., \& Bergeron, P. 2001, The Publications of the
  Astronomical Society of the Pacific, 113, 409

\bibitem[{Gagn{\'e} \& Faherty(2018)}]{2018ApJ...862..138G}
Gagn{\'e}, J., \& Faherty, J.~K. 2018, The Astrophysical Journal, 862, 138

\bibitem[{Gagn{\'e} {et~al.}(2015)Gagn{\'e}, Lafreni{\`e}re, Doyon, Malo, \&
  Artigau}]{2015ApJ...798...73G}
Gagn{\'e}, J., Lafreni{\`e}re, D., Doyon, R., Malo, L., \& Artigau, {\'E}.
  2015, The Astrophysical Journal, 798, 73

\bibitem[{Gagn{\'e} {et~al.}(2018)Gagn{\'e}, Roy-Loubier, Faherty, Doyon, \&
  Malo}]{Gagne:2018un}
Gagn{\'e}, J., Roy-Loubier, O., Faherty, J.~K., Doyon, R., \& Malo, L. 2018,
  arXiv.org, 1804.03093v1

\bibitem[{Gagn\'e {et~al.}(2018{\natexlab{a}})Gagn\'e, Mamajek, Malo, Riedel,
  Rodriguez, Lafreni\`~ere, Faherty, Roy-Loubier, Pueyo, Robin, \&
  Ren\'e}]{zenodobanyansigmaidl}
Gagn\'e, J., Mamajek, E.~E., Malo, L., {et~al.} 2018{\natexlab{a}}, {BANYAN
  $\Sigma$ (IDL) v1.1, Zenodo}, doi:10.5281/zenodo.1165086

\bibitem[{Gagn\'e {et~al.}(2018{\natexlab{b}})Gagn\'e, Mamajek, Malo, Riedel,
  Rodriguez, Lafreni\`~ere, Faherty, Roy-Loubier, Pueyo, Robin, \&
  Ren\'e}]{zenodobanyansigmapython}
---. 2018{\natexlab{b}}, {BANYAN $\Sigma$ (Python) v1.1, Zenodo},
  doi:10.5281/zenodo.1165085

\bibitem[{Gagn{\'e} {et~al.}(2018)Gagn{\'e}, Mamajek, Malo, Riedel, Rodriguez,
  Lafreni{\`e}re, Faherty, Roy-Loubier, Pueyo, Robin, \&
  Doyon}]{2018ApJ...856...23G}
Gagn{\'e}, J., Mamajek, E.~E., Malo, L., {et~al.} 2018, The Astrophysical
  Journal, 856, 23

\bibitem[{{Gaia Collaboration} {et~al.}(2018){Gaia Collaboration}, Brown,
  Vallenari, Prusti, de~Bruijne, \& {et al.}}]{GaiaCollaboration:2018io}
{Gaia Collaboration}, Brown, A. G.~A., Vallenari, A., {et~al.} 2018, Astronomy
  {\&} Astrophysics, doi:10.1051/0004-6361/201833051

\bibitem[{{Gaia Collaboration} {et~al.}(2016){Gaia Collaboration}, Prusti,
  de~Bruijne, Brown, Vallenari, Babusiaux, Bailer-Jones, Bastian, Biermann,
  Evans, Eyer, Jansen, Jordi, Klioner, Lammers, Lindegren, Luri, Mignard,
  Milligan, Panem, Poinsignon, Pourbaix, Randich, Sarri, Sartoretti, Siddiqui,
  Soubiran, Valette, van Leeuwen, Walton, Aerts, Arenou, Cropper, Drimmel,
  H{\o}g, Katz, Lattanzi, O'Mullane, Grebel, Holland, Huc, Passot, Bramante,
  Cacciari, Casta{\~n}eda, Chaoul, Cheek, De~Angeli, Fabricius, Guerra,
  Hern{\'a}ndez, Jean-Antoine-Piccolo, Masana, Messineo, Mowlavi, Nienartowicz,
  Ord{\'o}{\~n}ez-Blanco, Panuzzo, Portell, Richards, Riello, Seabroke, Tanga,
  Th{\'e}venin, Torra, Els, Gracia-Abril, Comoretto, Garcia-Reinaldos, Lock,
  Mercier, Altmann, Andrae, Astraatmadja, Bellas-Velidis, Benson, Berthier,
  Blomme, Busso, Carry, Cellino, Clementini, Cowell, Creevey, Cuypers,
  Davidson, De~Ridder, de~Torres, Delchambre, Dell'Oro, Ducourant, Fr{\'e}mat,
  Garc{\'\i}a-Torres, Gosset, Halbwachs, Hambly, Harrison, Hauser, Hestroffer,
  Hodgkin, Huckle, Hutton, Jasniewicz, Jordan, Kontizas, Korn, Lanzafame,
  Manteiga, Moitinho, Muinonen, Osinde, Pancino, Pauwels, Petit, Recio-Blanco,
  Robin, Sarro, Siopis, Smith, Smith, Sozzetti, Thuillot, van Reeven, Viala,
  Abbas, Abreu~Aramburu, Accart, Aguado, Allan, Allasia, Altavilla,
  {\'A}lvarez, Alves, Anderson, Andrei, Anglada~Varela, Antiche, Antoja, Anton,
  Arcay, Atzei, Ayache, Bach, Baker, Balaguer-N{\'u}{\~n}ez, Barache, Barata,
  Barbier, Barblan, Baroni, Barrado~y Navascu{\'e}s, Barros, Barstow, Becciani,
  Bellazzini, Bellei, Bello~Garc{\'\i}a, Belokurov, Bendjoya, Berihuete,
  Bianchi, Bienaym{\'e}, Billebaud, Blagorodnova, Blanco-Cuaresma, Boch,
  Bombrun, Borrachero, Bouquillon, Bourda, Bouy, Bragaglia, Breddels,
  Brouillet, Br{\"u}semeister, Bucciarelli, Budnik, Burgess, Burgon, Burlacu,
  Busonero, Buzzi, Caffau, Cambras, Campbell, Cancelliere, Cantat-Gaudin,
  Carlucci, Carrasco, Castellani, Charlot, Charnas, Charvet, Chassat,
  Chiavassa, Clotet, Cocozza, Collins, Collins, Costigan, Crifo, Cross, Crosta,
  Crowley, Dafonte, Damerdji, Dapergolas, David, David, De~Cat, de~Felice,
  de~Laverny, De~Luise, De~March, de~Martino, de~Souza, Debosscher, del Pozo,
  Delbo, Delgado, Delgado, di~Marco, Di~Matteo, Diakite, Distefano, Dolding,
  Dos~Anjos, Drazinos, Dur{\'a}n, Dzigan, Ecale, Edvardsson, Enke, Erdmann,
  Escolar, Espina, Evans, Eynard~Bontemps, Fabre, Fabrizio, Faigler,
  Falc{\~a}o, Farr{\`a}s~Casas, Faye, Federici, Fedorets,
  Fern{\'a}ndez-Hern{\'a}ndez, Fernique, Fienga, Figueras, Filippi, Findeisen,
  Fonti, Fouesneau, Fraile, Fraser, Fuchs, Furnell, Gai, Galleti, Galluccio,
  Garabato, Garc{\'\i}a-Sedano, Gar{\'e}, Garofalo, Garralda, Gavras, Gerssen,
  Geyer, Gilmore, Girona, Giuffrida, Gomes, Gonz{\'a}lez-Marcos,
  Gonz{\'a}lez-N{\'u}{\~n}ez, Gonz{\'a}lez-Vidal, Granvik, Guerrier, Guillout,
  Guiraud, G{\'u}rpide, Guti{\'e}rrez-S{\'a}nchez, {Guy, L. P.}, Haigron,
  Hatzidimitriou, Haywood, Heiter, Helmi, Hobbs, Hofmann, Holl, Holland, Hunt,
  Hypki, Icardi, Irwin, Jevardat~de Fombelle, Jofr{\'e}, Jonker, Jorissen,
  Julbe, Karampelas, Kochoska, Kohley, Kolenberg, Kontizas, Koposov,
  Kordopatis, Koubsky, Kowalczyk, Krone-Martins, Kudryashova, Kull, Bachchan,
  Lacoste-Seris, Lanza, Lavigne, Le~Poncin-Lafitte, Lebreton, Lebzelter,
  Leccia, Leclerc, Lecoeur-Taibi, Lemaitre, Lenhardt, Leroux, Liao, Licata,
  Lindstr{\o}m, Lister, Livanou, Lobel, L{\"o}ffler, L{\'o}pez, Lopez-Lozano,
  Lorenz, Loureiro, MacDonald, Magalh{\~a}es~Fernandes, Managau, Mann,
  Mantelet, Marchal, Marchant, Marconi, Marie, Marinoni, Marrese,
  Marschalk{\'o}, Marshall, Mart{\'\i}n-Fleitas, Martino, Mary, Matijevi{\v c},
  Mazeh, McMillan, Messina, Mestre, Michalik, Millar, Miranda, Molina,
  Molinaro, Molinaro, Moln{\'a}r, Moniez, Montegriffo, Monteiro, Mor, Mora,
  Morbidelli, Morel, Morgenthaler, Morley, Morris, Mulone, Muraveva, Musella,
  Narbonne, Nelemans, Nicastro, Noval, Ord{\'e}novic, Ordieres-Mer{\'e},
  Osborne, Pagani, Pagano, Pailler, Palacin, Palaversa, Parsons, Paulsen,
  Pecoraro, Pedrosa, Pentik{\"a}inen, Pereira, Pichon, Piersimoni, Pineau,
  Plachy, Plum, Poujoulet, Pr{\v s}a, Pulone, Ragaini, Rago, Rambaux,
  Ramos-Lerate, Ranalli, Rauw, Read, Regibo, Renk, Reyl{\'e}, Ribeiro,
  Rimoldini, Ripepi, Riva, Rixon, Roelens, Romero-G{\'o}mez, Rowell, Royer,
  Rudolph, Ruiz-Dern, Sadowski, Sagrist{\`a}~Sell{\'e}s, Sahlmann, Salgado,
  Salguero, Sarasso, Savietto, Schnorhk, Schultheis, Sciacca, Segol, Segovia,
  S{\'e}gransan, Serpell, Shih, Smareglia, Smart, Smith, Solano, Solitro,
  Sordo, Soria~Nieto, Souchay, Spagna, Spoto, Stampa, Steele,
  Steidelm{\"u}ller, Stephenson, Stoev, Suess, S{\"u}veges, Surdej, Szabados,
  Szegedi-Elek, Tapiador, Taris, Tauran, Taylor, Teixeira, Terrett, Tingley,
  Trager, Turon, Ulla, Utrilla, Valentini, van Elteren, Van~Hemelryck, van
  Leeuwen, Varadi, Vecchiato, Veljanoski, Via, Vicente, Vogt, Voss, Votruba,
  Voutsinas, Walmsley, Weiler, Weingrill, Werner, Wevers, Whitehead,
  Wyrzykowski, Yoldas, {\v Z}erjal, Zucker, Zurbach, Zwitter, Alecu, Allen,
  Allende~Prieto, Amorim, Anglada-Escude, Arsenijevic, Azaz, Balm, Beck,
  Bernstein, Bigot, Bijaoui, Blasco, Bonfigli, Bono, Boudreault, Bressan,
  Brown, Brunet, Bunclark, Buonanno, Butkevich, Carret, Carrion, Chemin,
  Ch{\'e}reau, Corcione, Darmigny, de~Boer, de~Teodoro, de~Zeeuw, Delle~Luche,
  Domingues, Dubath, Fodor, Fr{\'e}zouls, Fries, Fustes, Fyfe, Gallardo,
  Gallegos, Gardiol, Gebran, Gomboc, Gomez, Grux, Gueguen, Heyrovsky, Hoar,
  Iannicola, Isasi~Parache, Janotto, Joliet, Jonckheere, Keil, Kim, Klagyivik,
  Klar, Knude, Kochukhov, Kolka, Kos, Kutka, Lainey, LeBouquin, Liu, Loreggia,
  Makarov, Marseille, Martayan, Martinez-Rubi, Massart, Meynadier, Mignot,
  Munari, Nguyen, Nordlander, Ocvirk, O'Flaherty, Olias~Sanz, Ortiz, Osorio,
  Oszkiewicz, Ouzounis, Palmer, Park, Pasquato, Peltzer, Peralta, P{\'e}turaud,
  Pieniluoma, Pigozzi, Poels, Prat, Prod'homme, Raison, Rebordao, Risquez,
  Rocca-Volmerange, Rosen, Ruiz-Fuertes, Russo, Sembay, Serraller~Vizcaino,
  Short, Siebert, Silva, Sinachopoulos, Slezak, Soffel, Sosnowska, Strai{\v
  z}ys, ter Linden, Terrell, Theil, Tiede, Troisi, Tsalmantza, Tur, Vaccari,
  Vachier, Valles, Van~Hamme, Veltz, Virtanen, Wallut, Wichmann, Wilkinson,
  Ziaeepour, \& Zschocke}]{2016AA...595A...1G}
{Gaia Collaboration}, Prusti, T., de~Bruijne, J. H.~J., {et~al.} 2016,
  Astronomy and Astrophysics, 595, A1

\bibitem[{Garrison \& Gray(1994)}]{1994AJ....107.1556G}
Garrison, R.~F., \& Gray, R.~O. 1994, The Astronomical Journal, 107, 1556

\bibitem[{Glaspey(1971)}]{1971AJ.....76.1041G}
Glaspey, J.~W. 1971, Astronomical Journal, 76, 1041

\bibitem[{Gontcharov(2006)}]{2006AstL...32..759G}
Gontcharov, G.~A. 2006, Astronomy Letters, 32, 759

\bibitem[{Gutierrez-Moreno(1979)}]{1979PASP...91..299G}
Gutierrez-Moreno, A. 1979, Astronomical Society of the Pacific, 91, 299

\bibitem[{Hambly {et~al.}(2018)Hambly, Cropper, Boudreault, Crowley, Kohley,
  de~Bruijne, Dolding, Fabricius, \& {et al.}}]{Hambly:2018gr}
Hambly, N.~C., Cropper, M., Boudreault, S., {et~al.} 2018, Astronomy {\&}
  Astrophysics, doi:10.1051/0004-6361/201832716

\bibitem[{Hilton {et~al.}(2010)Hilton, West, Hawley, \&
  Kowalski}]{2010AJ....140.1402H}
Hilton, E.~J., West, A.~A., Hawley, S.~L., \& Kowalski, A.~F. 2010, The
  Astronomical Journal, 140, 1402

\bibitem[{H{\o}g {et~al.}(2000)H{\o}g, Fabricius, Makarov, Urban, Corbin,
  Wycoff, Bastian, Schwekendiek, \& Wicenec}]{2000AA...355L..27H}
H{\o}g, E., Fabricius, C., Makarov, V.~V., {et~al.} 2000, Astronomy and
  Astrophysics, 355, L27

\bibitem[{Holberg \& Bergeron(2006)}]{2006AJ....132.1221H}
Holberg, J.~B., \& Bergeron, P. 2006, The Astronomical Journal, 132, 1221

\bibitem[{Houk \& Cowley(1975)}]{1975mcts.book.....H}
Houk, N., \& Cowley, A.~P. 1975, University of Michigan Catalogue of
  two-dimensional spectral types for the HD stars. Volume I. Declinations -90
  to -53., I

\bibitem[{Kastner {et~al.}(2003)Kastner, Crigger, Rich, \&
  Weintraub}]{2003ApJ...585..878K}
Kastner, J.~H., Crigger, L., Rich, M., \& Weintraub, D.~A. 2003, The
  Astrophysical Journal, 585, 878

\bibitem[{Kirkpatrick {et~al.}(2011)Kirkpatrick, Cushing, Cruz, Gelino,
  Griffith, Skrutskie, Marsh, Wright, Mainzer, Eisenhardt, McLean, Thompson,
  Bauer, Benford, Bridge, Lake, Petty, Stanford, Tsai, Bailey, Beichman, Bloom,
  Bochanski, Burgasser, Capak, Hinz, Kartaltepe, Knox, Manohar, Masters,
  Morales-Calder{\'o}n, Prato, Rodigas, Salvato, Schurr, Scoville, Simcoe,
  Stapelfeldt, Stern, Stock, \& Vacca}]{2011ApJS..197...19K}
Kirkpatrick, D.~J., Cushing, M.~C., Cruz, K.~K., {et~al.} 2011, The
  Astrophysical Journal Supplement, 197, 19

\bibitem[{Kirkpatrick {et~al.}(2014)Kirkpatrick, Schneider, Fajardo-Acosta,
  Gelino, Mace, Wright, Logsdon, McLean, Cushing, Skrutskie, Eisenhardt, Stern,
  Balokovi{\'c}, Burgasser, Faherty, Lansbury, Rich, Skrzypek, Fowler, Cutri,
  Masci, Conrow, Grillmair, McCallon, Beichman, \& Marsh}]{2014ApJ...783..122K}
Kirkpatrick, D.~J., Schneider, A.~C., Fajardo-Acosta, S., {et~al.} 2014, The
  Astrophysical Journal, 783, 122

\bibitem[{Kowalski \& Saumon(2006)}]{2006ApJ...651L.137K}
Kowalski, P.~M., \& Saumon, D. 2006, The Astrophysical Journal, 651, L137

\bibitem[{Kuchner {et~al.}(2017)Kuchner, Faherty, Schneider, Meisner,
  Filippazzo, Gagn{\'e}, Trouille, Silverberg, Castro, Fletcher, Mokaev, \&
  Stajic}]{2017ApJ...841L..19K}
Kuchner, M.~J., Faherty, J.~K., Schneider, A.~C., {et~al.} 2017, The
  Astrophysical Journal Letters, 841, L19

\bibitem[{Lindegren {et~al.}(2018)Lindegren, Hern{\'a}ndez, Bombrun, Klioner,
  Bastian, \& Ramos-Lerate}]{Lindegren:2018gy}
Lindegren, L., Hern{\'a}ndez, J., Bombrun, A., {et~al.} 2018, Astronomy {\&}
  Astrophysics, doi:10.1051/0004-6361/201832727

\bibitem[{Lindegren {et~al.}(2016)Lindegren, Lammers, Bastian, Hern{\'a}ndez,
  Klioner, Hobbs, Bombrun, Michalik, Ramos-Lerate, Butkevich, Comoretto,
  Joliet, Holl, Hutton, Parsons, Steidelm{\"u}ller, Abbas, Altmann, Andrei,
  Anton, Bach, Barache, Becciani, Berthier, Bianchi, Biermann, Bouquillon,
  Bourda, Br{\"u}semeister, Bucciarelli, Busonero, Carlucci, Casta{\~n}eda,
  Charlot, Clotet, Crosta, Davidson, de~Felice, Drimmel, Fabricius, Fienga,
  Figueras, Fraile, Gai, Garralda, Geyer, Gonz{\'a}lez-Vidal, Guerra, Hambly,
  Hauser, Jordan, Lattanzi, Lenhardt, Liao, L{\"o}ffler, McMillan, Mignard,
  Mora, Morbidelli, Portell, Riva, Sarasso, Serraller, Siddiqui, Smart, Spagna,
  Stampa, Steele, Taris, Torra, van Reeven, Vecchiato, Zschocke, de~Bruijne,
  Gracia, Raison, Lister, Marchant, Messineo, Soffel, Osorio, de~Torres, \&
  O'Mullane}]{2016AA...595A...4L}
Lindegren, L., Lammers, U., Bastian, U., {et~al.} 2016, Astronomy {\&}
  Astrophysics, 595, A4

\bibitem[{Luri {et~al.}(2018)Luri, A~Brown, Sarro, Arenou, Bailer-Jones,
  Castro-Ginard, de~Bruijne, Prusti, Babusiaux, \& Delgado}]{Luri:2018eu}
Luri, X., A~Brown, A.~G., Sarro, L., {et~al.} 2018, Astronomy {\&}
  Astrophysics, doi:10.1051/0004-6361/201832964

\bibitem[{Malo {et~al.}(2014{\natexlab{a}})Malo, Artigau, Doyon,
  Lafreni{\`e}re, Albert, \& Gagn{\'e}}]{2014ApJ...788...81M}
Malo, L., Artigau, {\'E}., Doyon, R., {et~al.} 2014{\natexlab{a}}, The
  Astrophysical Journal, 788, 81

\bibitem[{Malo {et~al.}(2014{\natexlab{b}})Malo, Doyon, Feiden, Albert,
  Lafreni{\`e}re, Artigau, Gagn{\'e}, \& Riedel}]{2014ApJ...792...37M}
Malo, L., Doyon, R., Feiden, G.~A., {et~al.} 2014{\natexlab{b}}, The
  Astrophysical Journal, 792, 37

\bibitem[{Martin {et~al.}(2005)Martin, Fanson, Schiminovich, Morrissey,
  Friedman, Barlow, Conrow, Grange, Jelinsky, Milliard, Siegmund, Bianchi,
  Byun, Donas, Forster, Heckman, Lee, Madore, Malina, Neff, Rich, Small,
  Surber, Szalay, Welsh, \& Wyder}]{2005ApJ...619L...1M}
Martin, D.~C., Fanson, J., Schiminovich, D., {et~al.} 2005, The Astrophysical
  Journal, 619, L1

\bibitem[{Mermilliod(1987)}]{1987AAS...71..413M}
Mermilliod, J.~C. 1987, Astronomy and Astrophysics Supplement Series (ISSN
  0365-0138), 71, 413

\bibitem[{Mermilliod \& Mermilliod(1994)}]{1994cmud.book.....M}
Mermilliod, J.-C., \& Mermilliod, M. 1994, Catalogue of Mean UBV Data on Stars

\bibitem[{Messina {et~al.}(2010)Messina, Desidera, Turatto, Lanzafame, \&
  Guinan}]{2010AA...520A..15M}
Messina, S., Desidera, S., Turatto, M., Lanzafame, A.~C., \& Guinan, E.~F.
  2010, Astronomy and Astrophysics, 520, A15

\bibitem[{Mignard {et~al.}(2018)Mignard, Klioner, Lindegren, Hern{\'a}ndez,
  Bastian, Bombrun, Hobbs, Lammers, \& {et al.}}]{Mignard:2018bj}
Mignard, F., Klioner, S., Lindegren, L., {et~al.} 2018, Astronomy {\&}
  Astrophysics, doi:10.1051/0004-6361/201832916

\bibitem[{Nesterov {et~al.}(1995)Nesterov, Kuzmin, Ashimbaeva, Volchkov,
  R{\"o}ser, \& Bastian}]{1995AAS..110..367N}
Nesterov, V.~V., Kuzmin, A.~V., Ashimbaeva, N.~T., {et~al.} 1995, Astronomy and
  Astrophysics, 110, 367

\bibitem[{Ochsenbein {et~al.}(2000)Ochsenbein, Bauer, \&
  Marcout}]{2000AAS..143...23O}
Ochsenbein, F., Bauer, P., \& Marcout, J. 2000, Astronomy and Astrophysics
  Supplement, 143, 23

\bibitem[{Oelkers {et~al.}(2017)Oelkers, Stassun, \&
  Dhital}]{2017AJ....153..259O}
Oelkers, R.~J., Stassun, K.~G., \& Dhital, S. 2017, The Astronomical Journal,
  153, 259

\bibitem[{Oh {et~al.}(2017)Oh, Price-Whelan, Hogg, Morton, \&
  Spergel}]{2017AJ....153..257O}
Oh, S., Price-Whelan, A.~M., Hogg, D.~W., Morton, T.~D., \& Spergel, D.~N.
  2017, The Astronomical Journal, 153, 257

\bibitem[{Pecaut \& Mamajek(2013)}]{2013ApJS..208....9P}
Pecaut, M.~J., \& Mamajek, E.~E. 2013, The Astrophysical Journal Supplement,
  208, 9

\bibitem[{Perryman {et~al.}(1997)Perryman, Lindegren, Kovalevsky, Hoeg,
  Bastian, Bernacca, Cr{\'e}z{\'e}, Donati, Grenon, Grewing, van Leeuwen,
  van~der Marel, Mignard, Murray, Le~Poole, Schrijver, Turon, Arenou,
  Froeschl{\'e}, \& Petersen}]{1997AA...323L..49P}
Perryman, M. A.~C., Lindegren, L., Kovalevsky, J., {et~al.} 1997, Astronomy and
  Astrophysics 323, 323, L49

\bibitem[{Philip \& Egret(1980)}]{1980AAS...40..199P}
Philip, A.~D., \& Egret, D. 1980, Astronomy and Astrophysics Supplement, 40,
  199

\bibitem[{Platais {et~al.}(1998)Platais, Kozhurina-Platais, \& van
  Leeuwen}]{1998AJ....116.2423P}
Platais, I., Kozhurina-Platais, V., \& van Leeuwen, F. 1998, The Astronomical
  Journal, 116, 2423

\bibitem[{Reiners {et~al.}(2012)Reiners, Joshi, \&
  Goldman}]{2012AJ....143...93R}
Reiners, A., Joshi, N., \& Goldman, B. 2012, The Astronomical Journal, 143, 93

\bibitem[{Reis {et~al.}(2011)Reis, Corradi, de~Avillez, \&
  Santos}]{2011ApJ...734....8R}
Reis, W., Corradi, W., de~Avillez, M.~A., \& Santos, F.~P. 2011, The
  Astrophysical Journal, 734, 8

\bibitem[{Riaz {et~al.}(2006)Riaz, Gizis, \& Harvin}]{2006AJ....132..866R}
Riaz, B., Gizis, J.~E., \& Harvin, J. 2006, The Astronomical Journal, 132, 866

\bibitem[{Riello {et~al.}(2018)Riello, De~Angeli, Evans, Busso, Hambly,
  Davidson, Burgess, Montegriffo, Osborne, Kewley, Carrasco, Fabricius, Jordi,
  Cacciari, van Leeuwen, \& Holland}]{Riello:2018bo}
Riello, M., De~Angeli, F., Evans, D.~W., {et~al.} 2018, Astronomy {\&}
  Astrophysics, doi:10.1051/0004-6361/201832712

\bibitem[{Robert {et~al.}(2016)Robert, Gagn{\'e}, Artigau, Lafreni{\`e}re,
  Nadeau, Doyon, Malo, Albert, Simard, Bardalez~Gagliuffi, \&
  Burgasser}]{2016ApJ...830..144R}
Robert, J., Gagn{\'e}, J., Artigau, {\'E}., {et~al.} 2016, The Astrophysical
  Journal, 830, 144

\bibitem[{Rodriguez {et~al.}(2011)Rodriguez, Bessell, Zuckerman, \&
  Kastner}]{2011ApJ...727...62R}
Rodriguez, D., Bessell, M.~S., Zuckerman, B., \& Kastner, J.~H. 2011, The
  Astrophysical Journal, 727, 62

\bibitem[{Rodriguez {et~al.}(2013)Rodriguez, Zuckerman, Kastner, Bessell,
  Faherty, \& Murphy}]{2013ApJ...774..101R}
Rodriguez, D., Zuckerman, B., Kastner, J.~H., {et~al.} 2013, The Astrophysical
  Journal, 774, 101

\bibitem[{Sartoretti {et~al.}(2018)Sartoretti, Katz, Cropper, Panuzzo,
  Seabroke, Viala, Benson, Blomme, \& {et al.}}]{Sartoretti:2018jm}
Sartoretti, P., Katz, D., Cropper, M., {et~al.} 2018, Astronomy {\&}
  Astrophysics, doi:10.1051/0004-6361/201832836

\bibitem[{Schmidt {et~al.}(2007)Schmidt, Cruz, Bongiorno, Liebert, \&
  Reid}]{2007AJ....133.2258S}
Schmidt, S.~J., Cruz, K.~K., Bongiorno, B.~J., Liebert, J., \& Reid, N.~I.
  2007, The Astronomical Journal, 133, 2258

\bibitem[{Shaya \& Olling(2011)}]{2011ApJS..192....2S}
Shaya, E.~J., \& Olling, R.~P. 2011, The Astrophysical Journal Supplement, 192,
  2

\bibitem[{Skrutskie {et~al.}(2006)Skrutskie, Cutri, Stiening, Weinberg,
  Schneider, Carpenter, Beichman, Capps, Chester, Elias, Huchra, Liebert,
  Lonsdale, Monet, Price, Seitzer, Jarrett, Kirkpatrick, Gizis, Howard, Evans,
  Fowler, Fullmer, Hurt, Light, Kopan, Marsh, McCallon, Tam, Van~Dyk, \&
  Wheelock}]{2006AJ....131.1163S}
Skrutskie, M.~F., Cutri, R.~M., Stiening, R., {et~al.} 2006, The Astronomical
  Journal, 131, 1163

\bibitem[{Soubiran {et~al.}(2018)Soubiran, Jasniewicz, Chemin, Zurbach,
  Brouillet, Panuzzo, Sartoretti, Katz, \& {et al.}}]{Soubiran:2018fz}
Soubiran, C., Jasniewicz, G., Chemin, L., {et~al.} 2018, Astronomy {\&}
  Astrophysics, doi:10.1051/0004-6361/201832795

\bibitem[{Stauffer {et~al.}(2003)Stauffer, Jones, Backman, Hartmann, Barrado~y
  Navascu{\'e}s, Pinsonneault, Terndrup, \& Muench}]{2003AJ....126..833S}
Stauffer, J.~R., Jones, B.~F., Backman, D., {et~al.} 2003, The Astronomical
  Journal, 126, 833

\bibitem[{Taylor(2008)}]{2008AJ....136.1388T}
Taylor, B.~J. 2008, The Astronomical Journal, 136, 1388

\bibitem[{Torres {et~al.}(2008)Torres, Quast, Melo, \&
  Sterzik}]{2008hsf2.book..757T}
Torres, C. A.~O., Quast, G.~R., Melo, C. H.~F., \& Sterzik, M.~F. 2008,
  Handbook of Star Forming Regions, I, 757

\bibitem[{Tremblay {et~al.}(2011)Tremblay, Bergeron, \&
  Gianninas}]{2011ApJ...730..128T}
Tremblay, P.~E., Bergeron, P., \& Gianninas, A. 2011, The Astrophysical
  Journal, 730, 128

\bibitem[{van Leeuwen(2007)}]{2007AA...474..653V}
van Leeuwen, F. 2007, Astronomy {\&} Astrophysics, 474, 653

\bibitem[{Wright {et~al.}(2010)Wright, Eisenhardt, Mainzer, Ressler, Cutri,
  Jarrett, Kirkpatrick, Padgett, McMillan, Skrutskie, Stanford, Cohen, Walker,
  Mather, Leisawitz, Gautier, McLean, Benford, Lonsdale, Blain, Mendez, Irace,
  Duval, Liu, Royer, Heinrichsen, Howard, Shannon, Kendall, Walsh, Larsen,
  Cardon, Schick, Schwalm, Abid, Fabinsky, Naes, \& Tsai}]{2010AJ....140.1868W}
Wright, E.~L., Eisenhardt, P. R.~M., Mainzer, A.~K., {et~al.} 2010, The
  Astronomical Journal, 140, 1868

\bibitem[{Zuckerman {et~al.}(2006)Zuckerman, Bessell, Song, \&
  Kim}]{2006ApJ...649L.115Z}
Zuckerman, B., Bessell, M.~S., Song, I., \& Kim, S. 2006, The Astrophysical
  Journal, 649, L115

\bibitem[{Zuckerman \& Song(2004)}]{2004ARAA..42..685Z}
Zuckerman, B., \& Song, I. 2004, Annual Review of Astronomy {\&}Astrophysics,
  42, 685

\end{thebibliography}

\end{document}